\documentclass[preprint,1p]{elsarticle}
\usepackage[colorlinks,linkcolor=blue, citecolor=blue, urlcolor=blue]{hyperref}
\usepackage{graphicx}
\usepackage{dcolumn}
\usepackage{bm}
\usepackage{subfigure}
\usepackage{amsmath}
\usepackage{amssymb}
\usepackage{bbold}
\usepackage{bbm}
\usepackage{stmaryrd} 
\usepackage{geometry}
\usepackage{float}
\usepackage{multirow}
\usepackage{cuted}
\newcommand\vect[1]{\bm{#1}}

\newdefinition{rmk}{Remark}
\newproof{pf}{Proof}
\newproof{pot}{Proof of Theorem \ref{thm2}}

\begin{document}

\title{Fcc $\to$ bcc phase transition kinetics in an immiscible binary system: atomistic evidence of the twinning mechanism of transformation}%


\date{\today}


\author[1]{G. Demange}
\ead[1]{gilles.demange@univ-rouen.fr}
\author[1]{M. Lavrskyi}
\author[2]{K. Chen}%
\author[3]{X. Chen}%
\author[2,3]{Z. D. Wang}%
\ead[3]{wangzd@mater.ustb.edu.cn}
\author[1]{R. Patte}
\author[1]{H. Zapolsky}
\address[1]{GPM, UMR CNRS 6634, Universit\'e de Rouen-Normandy, 76575 Saint \'Etienne du Rouvray, France}%
\address[2]{School of Materials Science and Engineering, University of Science and Technology Beijing, Beijing 100083, P.R.China}
\address[3]{State Key Laboratory for Advanced Metals and Materials, University of Science and Technology Beijing, Beijing 100083, P.R.China}


\begin{abstract}

Extensive atomistic simulations based on the quasiparticle (QA) approach are performed to determine the momentous aspects of the displacive fcc/bcc phase transformation in a binary system. We demonstrate that the QA is able to predict the major structural characteristics of fcc/bcc phase transformations, including the growth of  a bcc nuclei  in a fcc matrix, and eventually the formation of an internally twinned structure consisting in two variants with  Kurdjumov-Sachs orientation relationship. At atomic level, we determine the defect structure of twinning boundaries and fcc/bcc interfaces, and identify the main mechanism for their propagation. In details, it is shown that twin boundaries are propagated by the propagation of screw dislocations in fcc along the $\langle\bar{1}\bar{1}1\rangle_{\alpha}$ direction, while the propagation of fcc screw dislocations along coherent terrace edges is the pivotal vector of the fcc/bcc transformation.  The simulation results are compared with our TEM and HRTEM observations of Fe-rich bcc twinned particle embedded in the fcc Cu-rich matrix in the Cu-Fe-Co system. 

\end{abstract}

\maketitle

\section*{Introduction}

Displacive solid-state phase transformations from the face centered cubic (fcc) austenite phase ($\gamma$)  to body centered cubic (bcc) ferrite  phase ($\alpha$) play a pivotal role in the physical properties of steels and ferrous alloys. It is characterized by a collective movement of a large number of atoms over a distance typically smaller than the interatomic distance. The rapid change in crystal structure inherently alters the mechanical properties of these materials, including fatigue, plasticity and strength, whence the early and thorough studies thereupon \cite{olson1976general,moritani2002comparison,sandvik1983characteristicsII,sandvik1983characteristicsIII,kelly1990orientation,qi2014microstructure,ma2007parent,bos2006molecular,ou2016molecular}.

The present understanding of fcc$\to$bcc transformation is based on the Phenomenological Theory of Martensite Crystallography (PTMC) \cite{wechsler1953theory,bowles1954crystallography}, which posits the existence of an invariant strain plane for the shape transformation \cite{wechsler1953theory,bowles1954crystallography}.  This is achieved by homogeneous deformations \cite{a1964introduction} giving rise to special Orientation Relationships (OR), and producing shape deformation manifesting themselves by a specific surface relief \cite{khachaturyan2013theory}. Aside from \cite{sandvik1983characteristicsIII}, the PTMC does not account for the atomic structure of the interface, and thus cannot explain the dislocation-based mechanisms fueling the propagation of the interface. This shortcoming was later addressed by the Topological Model (TM) \cite{pond2003comparison,ma2007parent,hirth2011compatibility}, which describes the structure of the fcc/bcc interface in terms of periodic unit of coherent terraces reticulated by a network of glissile transformation dislocations \cite{hirth1994dislocations,hirth2016disconnections}, and by the new theory of the fcc$\to$bcc transformation hereafter proposed in \cite{maresca2017austenite}.

In addition, numerical simulations such as molecular dynamics (MD) \cite{bos2006molecular,song2013atomistic} and Monte Carlo (MC) modeling \cite{castan1989kinetics,chen2015coupled}, have significantly contributed to unravel the atomic structure and propagation mode of the fcc/bcc interface. A distinct advantage of MD simulations in the study of interface migration is the fact that the motion of individual atoms can be monitored. The MD approach yet contends with several shortcomings. First, the growth mechanisms which can be derived for the MD simulations  \cite{ou2016molecular}, is strongly influenced  by the choice of interatomic potential \cite{johnson1989analytic,ackland1997computer,meyer1998martensite}. Second, albeit particularly suited to prospect the relaxation of fcc/bcc interfaces \cite{maresca2017austenite}, MD approaches are limited to reproduce the dynamics of fcc$\to$bcc transformations on large microstuctural units of several dozen nanometers, mainly due to computational limitations.  

An alternative  approach to MD is provided by the atomic phase-field class of models (APFM). Contrary to the standard phase-field method (PFM) \cite{chen2002phase}, which proved capable to reproduce the martensite transformation at mesoscale \cite{zhang2007phase,mamivand2013review}, the APFM operates at atomic space scale and diffusion time scale \cite{jin2006atomic,certain2011atomic}. Herein, it overrides the limitation of the standard PFM to a coarse grain description of the microstructure, while harnessing the numerical efficiency of phase-field algorithms \cite{demange2018generalization}. However, the APFM faces a host of difficulties, especially when connecting the atomic phase-field used in the model with the a real density and/or the position of atoms. This shortcoming hitherto precluded the accurate identification of salient features of the fcc$\to$bcc transformation, such as the atomic structure of twin boundaries (TB), fcc/bcc interface, and dislocations. Recently, a specific APFM coined quasi-particle (QA) was introduced in \cite{lavrskyi2016quasiparticle}. This approach extends the atomic density function theory proposed in \cite{khachaturyan2013theory} to the continuous case. It allows to circumvent some artifacts inherent in the original phase-field crystal (PFC) model \cite{elder2002modeling,elder2004modeling}. The QA was successful in modeling the structure of grain boundaries  in the bcc Iron phase \cite{kapikranian2014atomic,kapikranian2015point}, the self-assembly of atoms into complex structures \cite{lavrskyi2016quasiparticle}, as well as solute segregation in Fe-based alloys  \cite{mavrikakis2019multi}. 


The purpose of this study is to understand the dynamical process of the  fcc$\to$bcc transformation leading to a twin-like structure, from atomic to  microstructural level. The underlying motivation is twofold. First, it should demonstrate the potential of a fundamental atomic phase-field approach to recover some of the finest atomic characteristics of this class of phase transformations. Second, it should provide a unique opportunity to connect the atomistic mechanisms rooting the twinning fcc$\to$bcc transformation and the microstructure dynamics of the system. A first qualitative comparison between the numerical results and present experimental observations of internally twinned Fe-rich bcc particles having the Kurdjumov-Sachs (KS) orientation relationships (ORs) in a Fe-Cu-Co alloy \cite{chen_new} is also conducted, in order to challenge the model on a real study-case.

This study is organized as follows. First, the QA model for a binary system is introduced. Second, one large scale QA simulation of the fcc$\to$bcc transformation in a precipitate from a hypercritical bcc nuclei is presented as the cornerstone of the paper.  The morphology and the structure of the bcc inclusion are analyzed at microstructural scale.  Afterwards, twin boundaries are prospected, and the growth mode of twin variants is deciphered at atomic level. Finally, the surface relief of the fcc/bcc interface for an ellipsoidal bcc inclusion is characterized, whilst an interpretation of the propagation mechanism of the curved interface is proposed from the analysis of transformation dislocations. Simulation results are finally compared with the experimental data obtained by TEM observations on an as cast Fe-Cu-Co alloy.

\section{Numerical approach}

\subsection{Quasi-Particle model}
\label{QAmodel}

\begin{figure}[ht]
\centering
\subfigure[~~$\theta_{\alpha}(r)$]{\includegraphics[width=4cm]{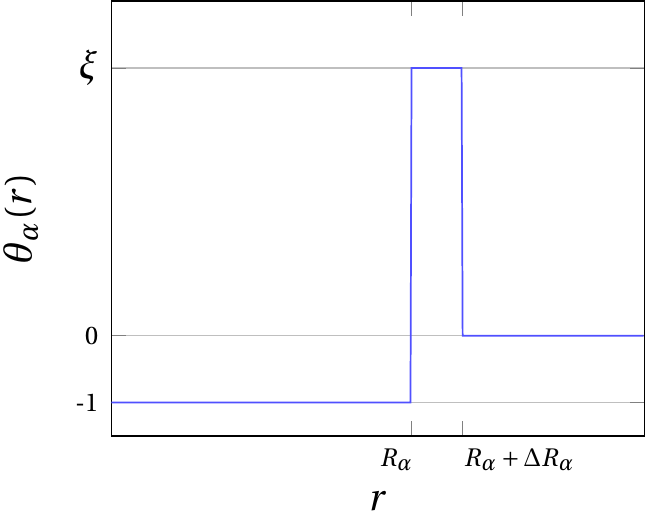}\label{fig:1a}}
\subfigure[~~$\hat{\theta}_{\alpha}(k)$]{\includegraphics[width=4cm]{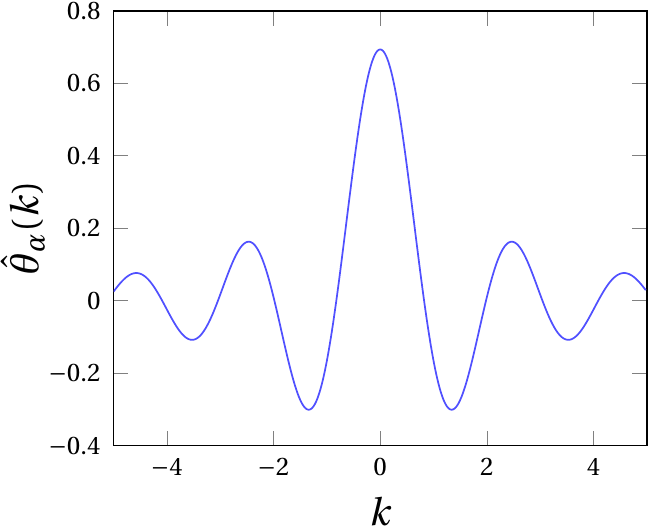}\label{fig:1b}}
\subfigure[~~$\hat{W}_{\alpha\beta}^{\text{LR}}$]{\includegraphics[width=4cm]{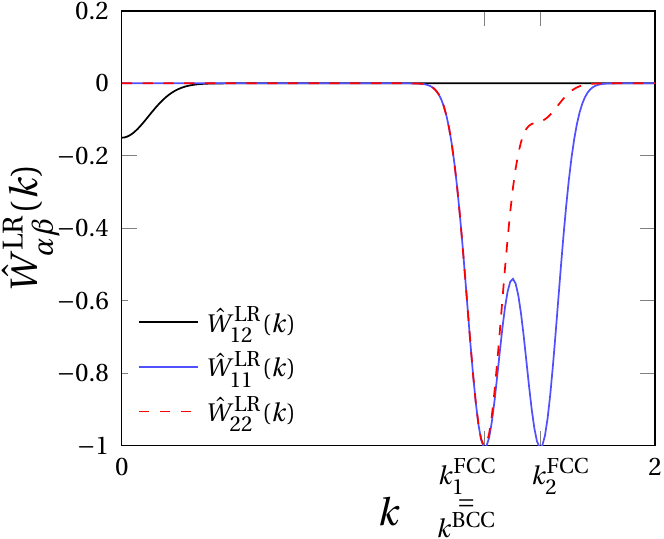}\label{fig:1c}}
\subfigure[~~$\hat{W}_{\alpha\beta}$]{\includegraphics[width=4cm]{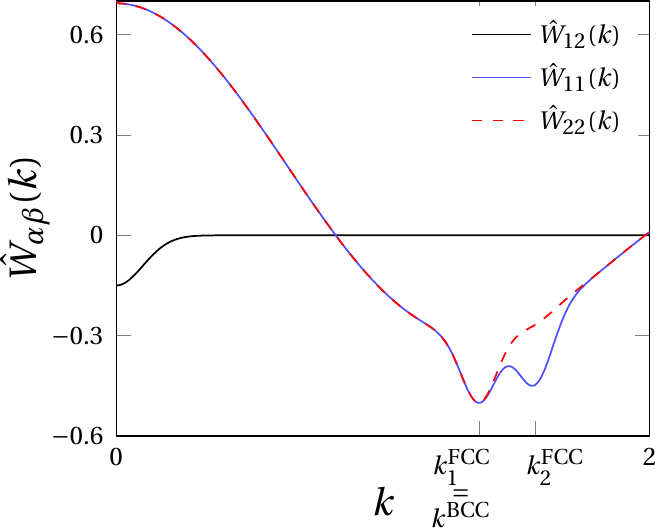}\label{fig:1d}}
\caption{Pairwise interaction potentials used in this work: (a) short range interaction potential in real space $\theta_{\alpha}(r)$ (for both components), (b) short range interaction potential in Fourier space $\hat{\theta}_{\alpha}(k)$, (c) long range interaction potentials $\hat{W}_{\alpha\beta}^{\text{LR}}$, (d) complete interaction potentials $\hat{W}_{\alpha\beta}$, using parameters $a_1=8$, $a_2=6.5$, $R_{\alpha}=2.81$, $\Delta R_{\alpha}=0.17$, and $\xi=4$,  $\sigma_1^{\text{fcc}}=\sigma_2^{\text{fcc}}=\sigma^{\text{bcc}}=\sigma^{12}=0.09$, $\lambda=0.2$, $\lambda_{12}=0.15$. In c) and d), red dashes: component 2 (bcc), blue: component 1 (fcc), black: chemical interactions between component 1 and component 2 atoms.}
\label{fig:1} 
\end{figure}

To describe the growth of the bcc particle in the fcc matrix in a binary system, the Quasi-Particle (QA) model was used \cite{lavrskyi2016quasiparticle}. Upon prescribing that the grid spacing $l_0$ of an Ising lattice $I=\left\{\vect{r}\right\}$ is several times smaller than the distance between neighboring atoms, QA assumes that each atom is a sphere comprising a certain number of Ising grid lattices. Lattices ascribed to atomic spheres are called fratons. Therefrom, the atomic configuration can be described by the occupation density function $\rho_{\alpha}(\vect{r},t)$,  defined as the ensemble average of fratonic configurations over a duration $t$ commensurate to a mean atomic diffusion time. This quantity can be interpreted as the probability to find a fraton in position $\vect{r}$ at time $t$. In the general case of a  $m$-components system, $m-1$ density probability functions $\left\{\rho_{\alpha}(\vect{r},t)\right\}_{\alpha=1,m-1}$ should be introduced, for the total  occupation density must be conserved ($\sum_{\alpha=1}^m \rho_{\alpha}(\vect{r},t)=1$. With these notations, the temporal evolution of the system is given by the microscopic diffusion equation \cite{jin2006atomic}:

\begin{equation}
\frac{\partial \rho_{\alpha}}{\partial t}(\vect{r},t)=\sum_{\beta=1}^m\sum_{\vect{r}'\in I}  L_{\alpha\beta} (\vect{r},\vect{r}')  \frac{\delta F}{\delta \rho_{\beta}(\vect{r}',t)}.
\label{eq1}
\end{equation}
Here, the $L_{\alpha,\beta}$ parameters are the kinetic coefficient matrices. They satisfy the condition $\sum_{\vect{r}\in I}L_{\alpha,\beta}(\vect{r})=0$ imposing the conservation of the average density  $\bar{\rho}_{\alpha}$. $F$ is the Helmholtz free energy written as:

\begin{equation}
F=\displaystyle\sum_{\alpha=1}^m \sum_{\substack{\beta=1\\\beta\geq \alpha}}^m \left[\frac{1}{2}\sum_{\vect{r},\vect{r}'\in I} W_{\alpha \beta}(\vect{r}-\vect{r}')\rho_{\alpha}(\vect{r},t)\rho_{\beta}(\vect{r}',t)\right]+k_bT\sum_{\vect{r}\in I} \left[\sum_{\alpha=1}^m\rho_{\alpha}\ln(\rho_{\alpha})+\left(1-\sum_{\alpha=1}^m\rho_{\alpha}\right)\ln\left(1-\sum_{\alpha=1}^m\rho_{\alpha}\right)\right].
\label{eq2}
\end{equation}
Therein, $W_{\alpha \beta}$ is a pairwise interaction potential, $k_b$ is the constant of Boltzmann, and $T$ is the temperature of the system. With this, the term on the left in equation \ref{eq2} corresponds to the internal energy, and the term on the right accounts for the configurational entropy. For convenience, the interaction potential $W_{\alpha \beta}(\vect{r})$ is implemented in reciprocal space via its Fourier transform  $\hat{W}_{\alpha \beta}(\vect{k})$, whereinto $\vect{k}$ is the k-vector defined by $\vect{k}=(k_x,k_y,k_z)=\frac{2\pi}{N}(h, k, l)$ with $(h, k, l)$ being dimensionless coordinates, and $N$ the size of the simulation box. The expression of $\hat{W}_{\alpha \beta}(\vect{k})$ is split into the so-called short range and long range interactions \cite{lavrskyi2016quasiparticle} given by:

\begin{equation}
\hat{W}_{\alpha \beta}(\vect{k})=\hat{\theta}_{\alpha}(\vect{k})\delta_{\alpha\beta}+\lambda \hat{W}_{\alpha \beta}^{\text{LR}}(\vect{k}),
\label{eq4}
\end{equation}
where the parameter $\lambda$ tunes the relative amplitude between short range and long range interactions. First, the short range interaction potential $\hat{\theta}_{\alpha}(\vect{k})$ can be seen as a hard sphere model for atoms. It is defined as the continuous Fourier  transform of the step function $\theta_{\alpha}(r)$ sketched in figure \ref{fig:1a}:

\begin{equation}
\begin{aligned}
\hat{\theta}_{\alpha}(k)&=\frac{4\pi}{k^3} \big[-\sin(kR_{\alpha}) +kR_{\alpha}\cos(kR_{\alpha})+\xi \big\{\sin(k(R_{\alpha}+\Delta R_{\alpha}))-k(R_{\alpha}+\Delta R_{\alpha})\cos(k(R_{\alpha}+\Delta R_{\alpha}))\\
&-\sin(kR_{\alpha})+kR_{\alpha}\cos(kR_{\alpha})\big\}\big],
\end{aligned}
\label{eq6}
\end{equation}
where $k=|\vect{k}|$, $R_{\alpha}$ is the radius of $\alpha$ atoms,  $\Delta R$ is the width of the repulsion part of the potential, and $\xi=|\max[\theta(r)]|/|\in[\theta(r)]|$. For each component, the long range interaction potential $\hat{W}_{\alpha \beta}^{\text{LR}}(\vect{k})$ stabilizes the desired crystal structure and fits both the elastic properties of the system and chemical interactions between atoms.

%

In this section, we propose a toy model of immiscible binary alloy ($m=2$), aiming at reproducing the displacive fcc$\to$bcc structural transition in a precipitate/matrix binary system. Component 1 will henceforth  pertain  to the matrix, and component 2 to the precipitate. In order to describe the fcc and bcc structures, the interaction potential $\hat{W}_{11}^{\text{LR}}(k)$ for the fcc structure was equipped with two wells of equal depth located in $k_1^{\text{fcc}}=\frac{2\pi\sqrt{3}}{a_1}$  and $k_2^{\text{fcc}}=\frac{4\pi}{a_1}$, versus one well centered in $k^{\text{bcc}}=\frac{2\pi\sqrt{2}}{a_2}$ for the bcc potential $\hat{W}_{22}^{\text{LR}}(k)$, where $a_1$ and $a_2$ are the lattice parameters of component 1 and 2 respectively. An additional well located in $k_2^{\text{fcc}}$ was adjoined to the potential $\hat{W}_{22}^{\text{LR}}(k)$, in order to stabilize the fcc structure far from the transition front between fcc and bcc structures. The amplitude of this well was mitigated by a factor $\epsilon<1$ to ensure that the bcc structure is more energetically favorable than the fcc structure.

The lattice parameters of the fcc and bcc structures were chosen as $a_1=8.0\Delta x$, $a_2=6.5\Delta x$ respectively (see table \ref{tab:1}), resulting in a very close position for the first well of the fcc structure, and the unique specific well of the bcc structure ($k_1^{\text{fcc}}\simeq k^{\text{bcc}}$). This tuning was used as a mean to keep the potential as simple as possible, as potential wells can herewith be found at two characteristic wavelength only ($k_1^{\text{fcc}}=\frac{2\pi\sqrt{3}}{a_1}$  and $k_2^{\text{fcc}}=\frac{4\pi}{a_1}$). Any other choice would have resulted in an additional well located in $k^{\text{bcc}} \neq k_1^{\text{fcc}}$, therefore requiring an enhanced numerical accuracy in the sampling of the Fourier space to be isolated one from another. Besides, the setting of the two lattice parameters as integer or half integer multiples of the grid lattice spacing ($a_1=8.0$, $a_2=6.5$) was imposed by the sensibility of the the QA model in real space.  This choice comes with a small but non zero lattice misfit $\delta_{\text{fcc/bcc}}=0.0049$. Enhanced accuracy can be achieved in the QA approach, provided more grid spacing are used to span crystal lattice parameters, yet at the cost of a smaller simulation domain. Now, the present study required to use the largest numerically reachable spatial domain.

In this study, only spherically symmetric potentials were used as a mean to allow the formation of local clusters of different metastable phases with any orientations at the interface. Choice was made to use linear combinations of Gaussian functions. To mimic the immiscibility of components 1 and 2, a repulsive cross interaction potential  $\hat{W}_{12}^{\text{LR}}(k)$  was added, which can be related to the mixing energy of the binary system. It is defined as a simple Gaussian function centered in $k=0$, allowing the phase separation of the different chemical species.  Then, the corresponding long range interaction potentials read:

\begin{equation}
\left\{
\begin{aligned}
&\hat{W}_{11}^{\text{LR}}(k)=-\exp\left(-\frac{(k-k_1^{\text{fcc}})^2}{2(\sigma_1^{\text{fcc}})^2}\right)-\exp\left(-\frac{(k-k_2^{\text{fcc}})^2}{2(\sigma_2^{\text{fcc}})^2}\right)\\
&\hat{W}_{22}^{\text{LR}}(k)=-\exp\left(-\frac{(k-k^{\text{bcc}})^2}{2(\sigma^{\text{bcc}})^2}\right)-\epsilon\exp\left(-\frac{(k-k_2^{\text{fcc}})^2}{2(\sigma_2^{\text{fcc}})^2}\right)\\
&\hat{W}_{12}^{\text{LR}}(k)=-\lambda_{12}\exp\left(-\frac{k^2}{2(\sigma^{12})^2}\right),
\end{aligned}
\right.
\label{eq7}
\end{equation}
where the tuning of the standard deviations $\sigma_1^{\text{fcc}}$, $\sigma_2^{\text{fcc}}$, and $\sigma^{\text{bcc}}$ of the Gaussian functions modulates the elastic properties of the material  \cite{lavrskyi2017modelisation,vaugeois2017modelisation}. Rather than the absolute values of elastic constants in reduced units, the yield between the bulk modulus $B^{\text{bcc}}$ and $B^{\text{fcc}}$ of each phase, as well as Zener anisotropy parameters $A^{\text{bcc}}$ and $A^{\text{fcc}}$ have a significant influence on the transformation. Following the procedure for elastic constants determination proposed in a previous work \cite{vaugeois2017modelisation}, the parameter setting provided in table \ref{tab:1} provides $A^{\text{bcc}}=1.45$ and $A^{\text{fcc}}=1.05$ (elastically soft directions are of $\langle 100\rangle_{\gamma}$ type), and  $B^{\text{bcc}}/B^{\text{fcc}}=1.41$ (the bcc structure is harder than the fcc structure). A qualitative comparison between QA simulations and experiments in a Fe-Cu-Co alloy displaying twinned Iron-rich bcc precipitates in a Copper fcc matrix is proposed in the last section of this study. While the elastic constants used in the QA are not fitted on their experimental counterpart in the Fe-Cu-Co alloy, Zener anisotropy parameters for the bcc and fcc structures in both the QA model and experiments are nonetheless greater than 1.0. Moreover, the yield between the bulk modulus of the bcc and fcc structures in the QA (1.41) is tantamount to the experimental values for bcc Fe/fcc Fe (1.27) and bcc Fe/fcc Cu (1.35) \cite{zarestky1987lattice,lide2004crc}. Finally, the parameter $\lambda_{12}$ weights the relative influence of the structural contributions $\hat{W}_{11}^{\text{LR}}(k)$ $\hat{W}_{22}^{\text{LR}}(k)$ with respect to the chemical repulsion $\hat{W}_{12}^{\text{LR}}(k)$. The full interaction potential \ref{eq4} is displayed in figure \ref{fig:1d}.


Simulations were performed in reduced units. The average density of probability $\bar{\rho}_{1,2}$ of type 1 and 2 atoms was defined as $(4\pi R_{1,2}^3N_{1,2}/(3V)$, where $V$ is the total volume of the system, and $N_{1,2}$ is the total number of atoms of type 1 and 2 at ground state. The input parameter in simulations was thus $\bar{\rho}= \bar{\rho}_{1}+\bar{\rho}_{2}$. Moreover, $k_bT$ and $\xi$ were expressed in $k_bT_m$ units, where $T_m$ is the melting temperature of the system with composition $(\bar{\rho}_{1},\bar{\rho}_{2})$. The space scale was chosen as the grid spacing $l_0$, as set by the number of grid lattices spanning one lattice parameter. The associated time scale is $t_0=l_0^2/(Mk_bT_m)$ where  $M$ is an average mobility. Considering the optimized computational features of Fourier based numerical schemes, the kinetic equation \ref{eq1} for the density probability functions $\rho_{1,2}$ was solved in Fourier space:

\begin{equation}
\left\{
\begin{aligned}
\frac{\partial \hat{\rho}_1}{\partial t}(\vect{k},t)&=  \hat{\bar{L}}_{11} (k)\left[\hat{W}_{11}(k)\hat{\rho}_1(\vect{k},t)+ \hat{W}_{12}(k)\hat{\rho}_2(\vect{k},t) +k_bT\left\{\ln\big(\rho_1/(1-\rho_1-\rho_2)\big)  \right\}_{\vect{k}} \right]\\
&+ \hat{\bar{L}}_{12} (k)\left[ \hat{W}_{22}(k)\hat{\rho}_2(\vect{k},t) +\hat{W}_{12}(k)\hat{\rho}_1(\vect{k},t)+k_bT\left\{\ln\big(\rho_2/(1-\rho_1-\rho_2)\big)  \right\}_{\vect{k}} \right]\\
\frac{\partial \hat{\rho}_2}{\partial t}(\vect{k},t)&= \hat{\bar{L}}_{22} (k)\left[  \hat{W}_{22}(k)\hat{\rho}_2(\vect{k},t) +\hat{W}_{12}(k)\hat{\rho}_1(\vect{k},t) +k_bT\left\{\ln\big(\rho_2/(1-\rho_1-\rho_2)\big)  \right\}_{\vect{k}}\right]\\
&+ \hat{\bar{L}}_{12} (k)\left[\hat{W}_{11}(k)\hat{\rho}_1(\vect{k},t)+ \hat{W}_{12}(k)\hat{\rho}_2(\vect{k},t) +k_bT\left\{\ln\big(\rho_1/(1-\rho_1-\rho_2)\big)  \right\}_{\vect{k}}\right],\\
\end{aligned}
\right.
\label{eq9}
\end{equation}
where $\hat{\rho}_{\alpha}$ is the Fourier transform of the fraton density function $\rho_{\alpha}$, $\hat{\bar{L}}_{\alpha\beta} (k)=-\bar{M}_{\alpha\beta}k^2$ are the reduced kinetic coefficient matrices in Fourier space under spherical invariance hypothesis for isotropic diffusion, and $\left\{\cdot \right\}_{\vect{k}}$ is the discrete Fourier transform operator. All parameters are compiled  in reduced units  in table \ref{tab:1}.

\begin{table}[h]
	\begin{center}
		\begin{tabular}{|llllllllllllll|}
			\hline
			$a_{1,2}$  & $R_{1,2}$ &  $\Delta R_{1,2}$ & $\xi$ & $\lambda$ & $\epsilon$ & $\sigma_{1,2}^{\text{fcc}}$  & $\sigma^{\text{bcc}}$  &  $\sigma^{\text{12}}$ & $\lambda^{12}$ & $k_bT$ & $\bar{\rho}$ & $\bar{M}_{1,2}$ & $\bar{M}_{12}$ \\
			\hline
			 8.0/6.5  & 2.81 &  0.17 &   4.0 &   0.2 & 0.1 &  0.09 &   0.09 &  0.09 &      0.15  &  0.11 & 0.13 & 1.0 & 1.0\\
			\hline
		\end{tabular}
		\caption{Parameters in formulas \ref{eq4}, \ref{eq6} and \ref{eq7} used in simulations.}
		\label{tab:1}
	\end{center}
\end{table}

To model the fcc to bcc phase transformation in the bosom of the precipitate, the initial condition was chosen as a pure component 2 particle with perfect cubic shape embedded in a pure component 1 matrix with cube to cube OR. A small bcc nuclei was included at the center of the precipitate with KS OR with the fcc phase. This initial condition was used as a mean to overcome the nucleation barrier of the bcc nucleus, inasmuch as the nucleation step of the process remains beyond the scope of this study. The OR of the bcc nuclei corresponded to the first Kurdjumov-Sachs (KS) variant $V_1$. The corresponding rotation matrix is provided in the  appendix  \ref{app}.
In this work,  the length scale of the model was set to $l_0\simeq 0.045$ nm corresponding to the Fe-Cu alloy ($a_1\sim 0.36$). In contrast, all simulations were performed in reduced time units considering the complex dependence of the time scale on the thermodynamic and kinetic parameters of the system. Simulations were performed in three dimensions on a $1024^3$ grid lattice equipped with periodic boundary conditions. For the chosen length scale ($l_0\simeq 0.045$ nm), this corresponds to a volume of ($46$ nm)$^3$. The kinetic equations \ref{eq9} were solved by the Spectral-Eyre scheme \cite{demange2018generalization}, with the reduced time step $\Delta t=0.005$, on 512 cores of the supercalculator CRIANN of Normandy.

QA simulations were eventually post-treated to accurately spot atom positions. Indeed, the QA  is intrinsically a continuous approach. At finite  temperature, atoms are associated to atomic spheres with the fraton density function profile resembling a Gaussian. However, numerical fluctuations of the atomic density fields $\rho_{\alpha}(\vect{r},t)$, as well as the emergence of partially delocalized, splitting, and coalescing atoms at interfaces (grain boundaries, fcc/bcc interface) and within the bulk (dislocation propagation) spices things up to accurately spot the center of atoms. While this behavior grants the method with its physical versatility and computational efficiency, it is a true hurdle when it comes to characterize defect structures. The \texttt{fratons2atoms}  package \cite{fratons2atoms} was harnessed using default parameters to interpret the structures from QA calculations and make an educated guess of the most reliable deterministic atomic structure associated to a given QA simulation.

\section{Numerical results}

\subsection{Kinetic process of the fcc$\to$bcc transformation}

\begin{figure}[ht]
\centering
\includegraphics[width=16cm]{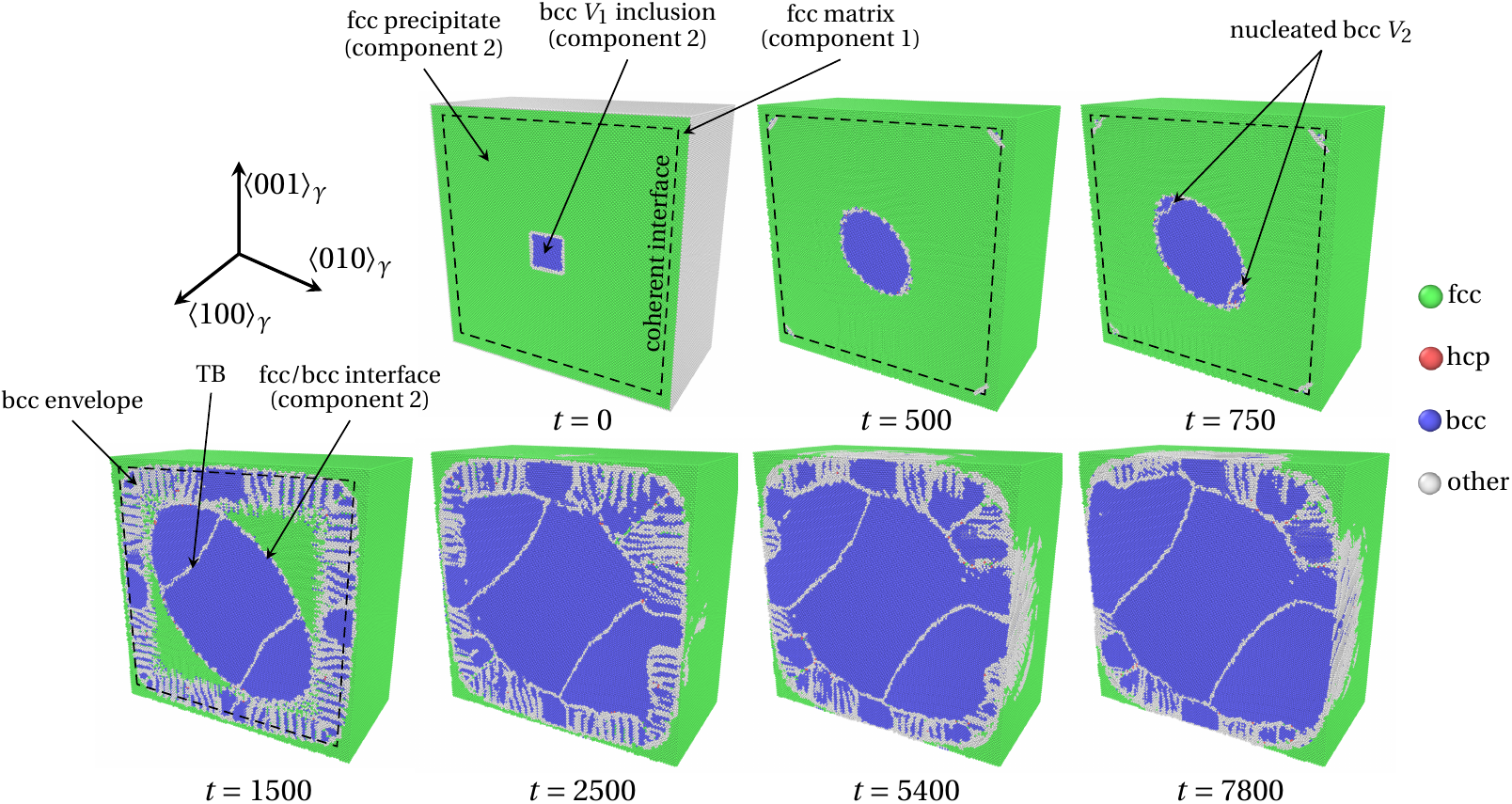}
\caption{Kinetic evolution of the precipitate structure (component 2), as simulated in the QA model using numerical parameters listed in table \ref{tab:1}, starting from an initial bcc nuclei with KS OR ($V_1$). Visualization via the  CNA of OVITO, after extraction of atom centers from the atomic density field $\rho_2$, using \texttt{fratons2atoms}, on a 800$^3$ visualization subdomain of the 1024$^3$ simulation domain. Green: fcc, blue: bcc, red: hcp and grey: unknown (perturbed structure). The bcc $V_1$ inclusion grows in a direction close to $\langle 1\bar{2}1\rangle_{\gamma}$ within the fcc structure until full fcc$\to$bcc transformation is achieved in the precipitate. Two new bcc structure domains are formed, and a highly perturbed bcc envelope appears.}
\label{fig:2} 
\end{figure}

In this work, main attention was put on the different stages of the shape evolution and atomic structure of the bcc inclusion in the fcc matrix. This is displayed in \ref{fig:2}. However, the evolution of the matrix and the precipitate-matrix interface was not considered. In figure \ref{fig:2}, the Common Neighbor Analysis (CNA) from OVITO was used, where green is for the fcc ($\gamma$) structure, blue for bcc ($\alpha$), and red for hcp. Any other crystallography is indicated in gray. This corresponds to  perturbed (non crystalline) structures, including interfaces and boundaries. One should remain cautious here, as only a 800$^3$ visualization subdomain of the 1024$^3$ simulation domain, which focuses on the precipitate is displayed, so that a significant portion of the fcc matrix (component 1, outside the region circumscribed by the dashed line) does not appear.

\begin{figure}[ht]
\centering
\includegraphics[height=5.5cm]{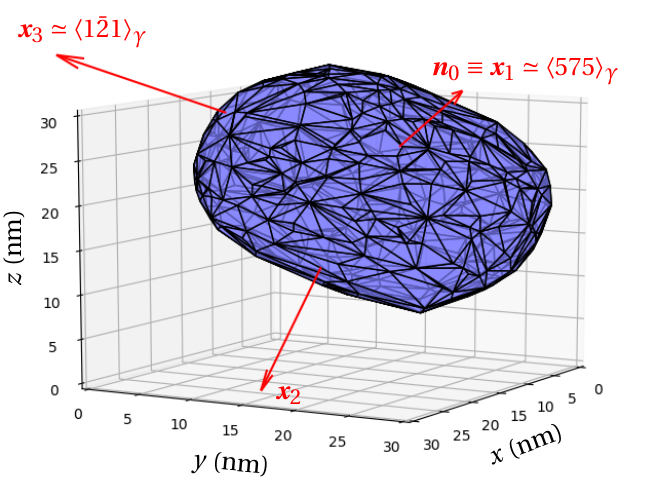}
\caption{Convex hull of the bcc inclusion at $t=1500$ and 3 principal axis $\vect{x}_1$ ($\sim$ invariant plane strain direction $\vect{n}_0$), $\vect{x}_2$ and $\vect{x}_3$ ($\sim$ preferential growth direction) of the fitting ellipsoid, sorted in increasing order of ellipsoid parameter.}
\label{fig:3} 
\end{figure}

The different steps of the bcc inclusion growth are shown in figure \ref{fig:2}. It grows with a roughly ellipsoidal morphology, flattened in a direction $\vect{n}_0$. Based on Eshelby's theory of coherent inclusion \cite{khachaturyan2013theory,eshelby1957determination}, a plate like morphology whose habit plane of normal vector  $\vect{n}_0$ aligns with the invariant plane is prone to minimize the bulk strain energy of the inclusion. Along this line,  the flattened direction $\vect{n}_0$ of the inclusion can be used as a first yardstick of the invariant plane direction normal. To estimate $\vect{n}_0$, the convex hull of the the bcc inclusion  was calculated, and the least square ellipsoidal fitting of this surface was performed upon extending the constrained minimization method proposed in \cite{fitzgibbon1996m}, from two to three dimensions. The convex hull of the bcc inclusion at $t=1500$ is displayed in figure \ref{fig:3}. It is equipped with the principal axis $\vect{x}_1$, $\vect{x}_2$ and $\vect{x}_3$ of the fitting ellipsoid, sorted in increasing order of ellipsoid parameter. It was found that $\vect{n}_0\equiv\vect{x}_1=\langle 0.56,0.67,0.49\rangle$, which forks off the direction  normal to the $(575)_{\gamma}$ plane by $3.8^{\circ}$ only. The latter is a usual habit plane of a  martensite phase embedded in fcc austenite phase in many steels and Iron based alloys, and more specifically in lath martensites  \cite{sandvik1983characteristicsII,qi2014microstructure,krauss2015steels,nishiyama2012martensitic,sandvik1983characteristicsI,kelly1992crystallography}.  Besides, $\vect{x}_3$ corresponds to the largest ellipsoid parameter, and can thus be interpreted as the  the preferential growth direction. It was found that $\vect{x}_3=\langle 0.52,-0.74,0.42\rangle$, which draws near to the $\langle 1\bar{2}1\rangle_{\gamma}$ crystal orientation ($7.57^{\circ}$ deviation).

In figure \ref{fig:2}, new structural domain siding the inclusion are formed with a different crystal orientations when the size of the initial variant exceeds a critical value ($\sim 13-14$ nm), roughly reached at $t=750$. A twin-like structure emerges therefrom. In parallel, an envelope with a perturbed structure is formed at the boundaries of the precipitate between fcc and bcc phases for $t>1500$. This secondary transformation is triggered by a consequent strain at the precipitate/matrix interface (not shown). This strain stems from the chemical repulsion between atoms of different species at the precipitate/matrix interface. After some time, the envelope  decomposes into numerous thin bcc structure domains ($t=1500$). At longer times, most structure domains coalesce, whilst the envelope itself merges with the bcc inclusion ($t=7800$). It is likely that the formation of this bcc envelope is a transient process, which allows the system to relax the stress at the precipitate/matrix interface.

\subsection{Twinning structure of the bcc inclusion}

\begin{figure}[ht]
\centering
\includegraphics[height=7.5cm]{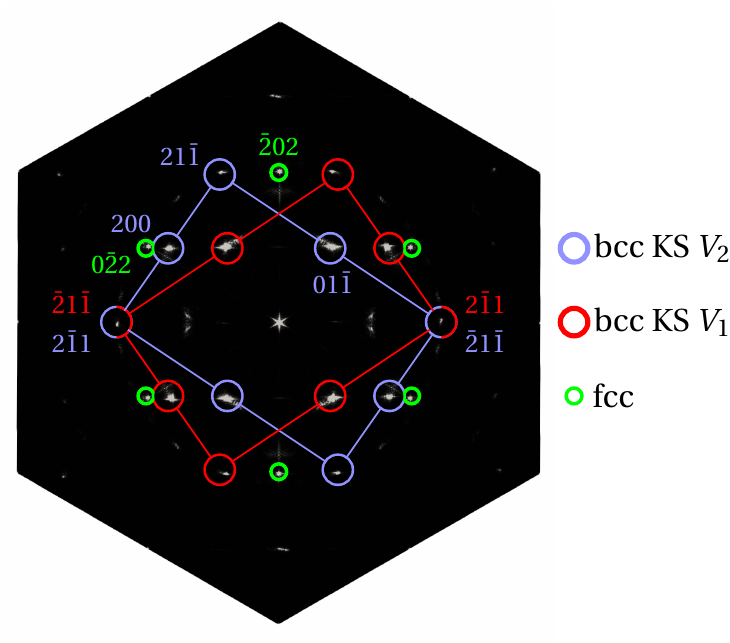}
\caption{Diffraction pattern in the $(111)_{\gamma}$ plane, as obtained from the projection of the diffraction intensity $I_2(\vect{k})$ of the second component (precipitate). The diffraction spots corresponding to the bcc structure with KS OR $V_1$ are indicated in red, KS OR $V_2$ in blue, and fcc in green.}
\label{fig:4} 
\end{figure} 	

\paragraph{Diffraction analysis of the twinning structure}

The identification of the crystallography of the structure domains within the inclusion was achieved via the diffraction analysis of atomic configurations extracted from QA simulations. To that end, the diffraction intensity $I_2(\vect{k})=\hat{\rho}_2(\vect{k},t)\hat{\rho}_2^*(\vect{k},t)$ of the component 2 precipitate was calculated, where $\hat{\rho}_2^*$ refers to the complex conjugate of $\hat{\rho}_2$. The simulated diffraction pattern in the $(111)_{\gamma}$ plane at $t=1500$ is presented  in figure \ref{fig:4}. Diffraction spots corresponding to three crystal structures can be observed. Each structure is identified by different color circles. Green circles single out the diffraction  spots of the fcc structure, including  $(\bar{2}02)_{\gamma}$ and $(0\bar{2}2)_{\gamma}$ points. Red circles enclose the diffraction spots of the bcc structure with KS OR $V_1$ ($\alpha_1$ phase),   such as  $(2\bar{1}1)_{\alpha_1}$ and  $(\bar{2}1\bar{1})_{\alpha_1}$. Finally, the diffraction spots emphasized by blue circles can be obtained by reflection of the diffraction pattern of the KS OR $V_1$ variant. The superimposition of two (blue) spots with the  $(2\bar{1}1)_{\alpha_1}$ and $(\bar{2}1\bar{1})_{\alpha_1}$ (red) spots additionally  indicates that the  structure corresponding to the blue spots is the mirror image of the bcc structure with KS OR $V_1$  in the $(2\bar{1}1)_{\alpha_1}$ plane. This is the bcc structure with KS OR $V_2$  ($\alpha_2$ phase, see appendix \ref{app}). Therefrom, we deduce that the structure domains formed after $t=750$ have the $V_2$ KS OR.

The formalism of deformation twinning \cite{christian1995deformation} provides a convenient albeit simplified toolbox to characterize the present twinning structure. Based on the analysis of figure \ref{fig:4}, the twinning plane is $K_1=(2\bar{1}1)_{\alpha_1}$, and the shear plane is $P=(111)_{\gamma}\parallel (011)_{\alpha_1}$. The twinning direction can then be defined as the intersection line between the shear plane and the twinning plane, namely $\eta_1=\langle\bar{1}\bar{1}1\rangle_{\alpha_1}$. The corresponding twinning mode is thus $(2\bar{1}1)_{\alpha_1}|\langle\bar{1}\bar{1}1\rangle_{\alpha_1}$. It is prevalent in various Iron based alloys \cite{monzen1992face,le1993effects}. In addition, it was found that the misorientation angle $\theta\simeq 69.5^{\circ}$ between twinning variants $V_1$ and $V_2$ around the rotation axis $\langle 011\rangle_{\alpha_1}$ was consistent with the predicted value  $\theta^{\text{id}}= 70.5^{\circ}$. For this mode, twinning proceeds by homogeneous simple shear deformation of amplitude $s=1/\sqrt{2}$, without shuffling. This deformation achieves perfect coincidence of lattices of both twins in the shear plane.


\begin{figure}[ht]
\centering
\subfigure[~~IR+RI couple]{\includegraphics[width=7cm]{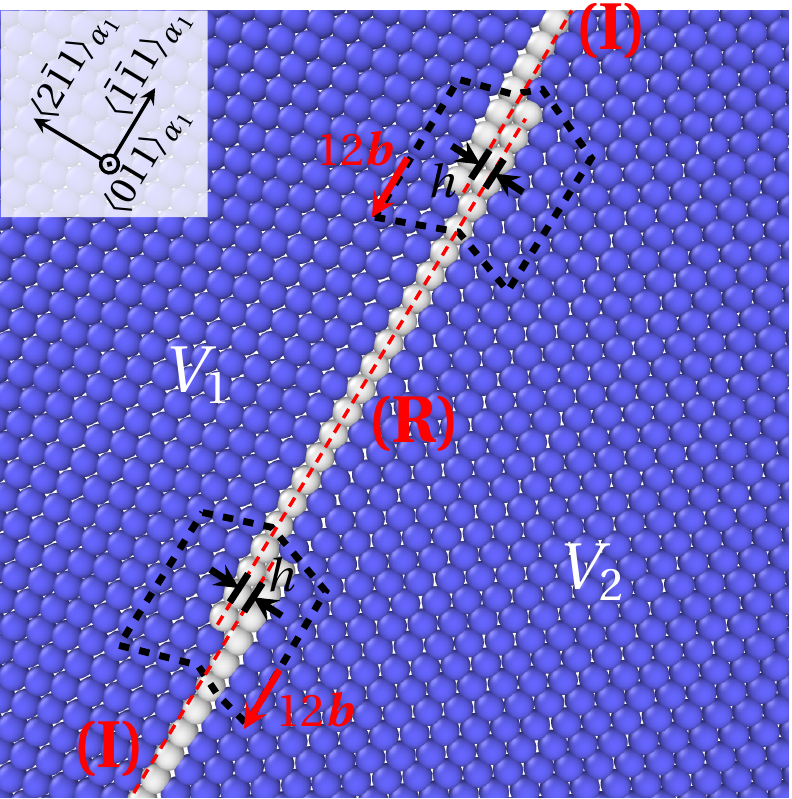}\label{fig:5a}}\hspace{1cm}
\subfigure[~~Z dislocation (1)]{\includegraphics[width=7cm]{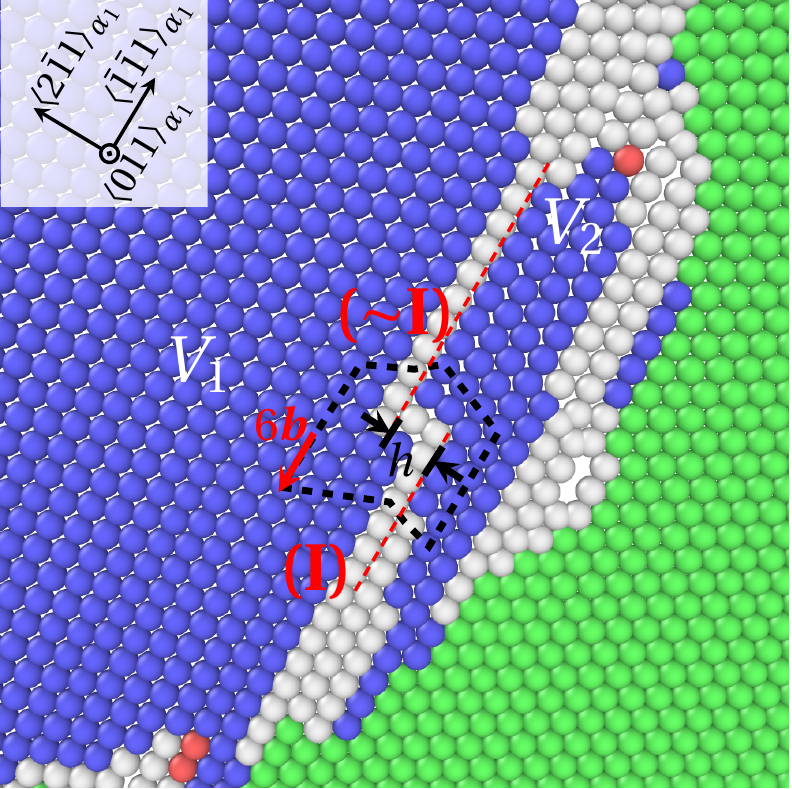}\label{fig:5b}}
\caption{Twin dislocations analysis in the close packed plane $(111)_{\gamma}$ from QA simulations at $t=1500$. (a) Partial twin dislocation couple (IR+RI) with step $h=a_2/(2\sqrt{6})$, Burgers vector $\vect{b}=\frac{1}{12}\langle\bar{1}\bar{1}1\rangle_{\alpha_1}$, connected by a strip of reflection TB (R). Elsewhere, the TB is isosceles (I). (b) Zonal twin dislocation (Z) with step $h=2a_2/\sqrt{6}$, Burgers vector $\vect{b}=\frac{1}{6}\langle\bar{1}\bar{1}1\rangle_{\alpha_1}$. A video of partial twin dislocations motion can be found in the Online Supplementary Material. }
\label{fig:5} 
\end{figure} 	

\paragraph{Twin boundaries (TB) and twin dislocations}

\begin{figure}[ht]
\centering
\subfigure[~~(I) and (R) twin boundary (theory)]{\includegraphics[width=12cm]{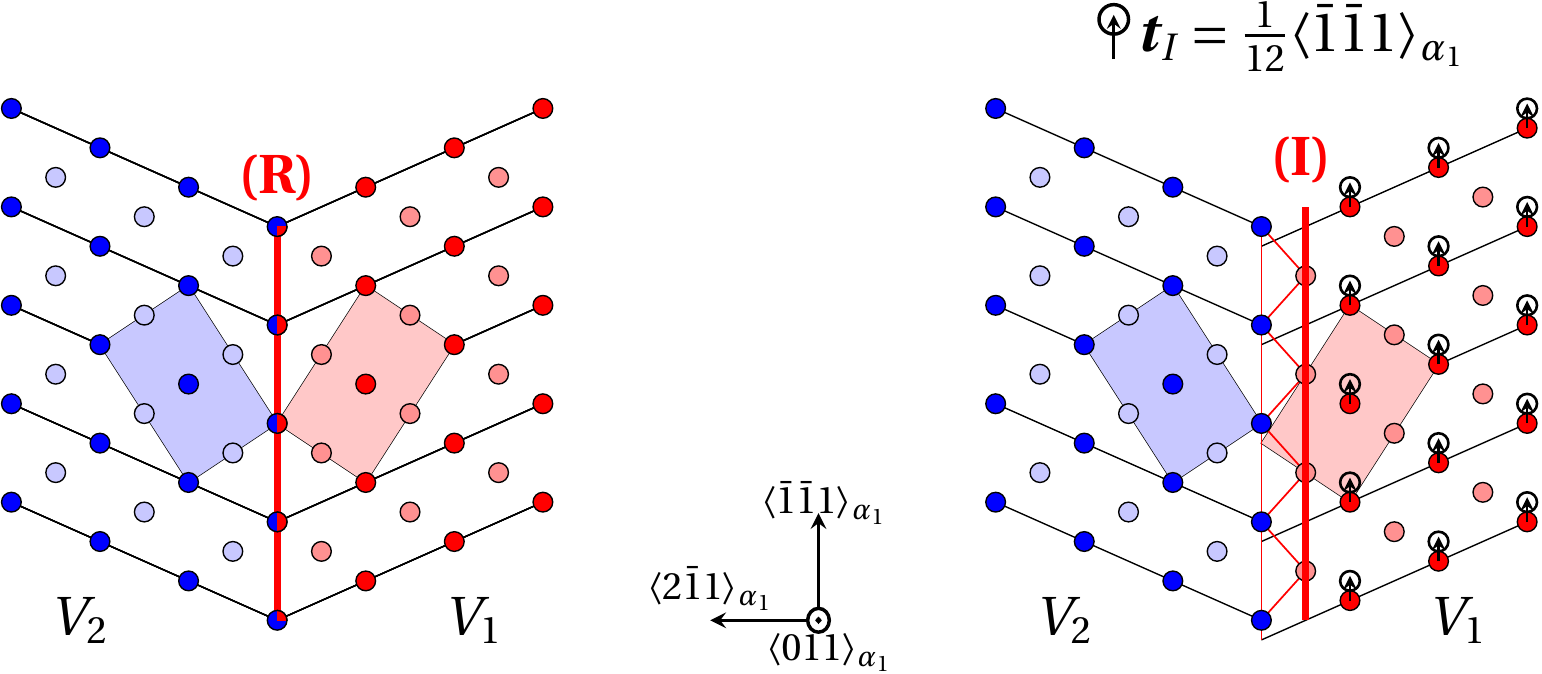}\label{fig:6a}}
\subfigure[~~(I) and (R) twin boundary (simulation)]{\includegraphics[width=12cm]{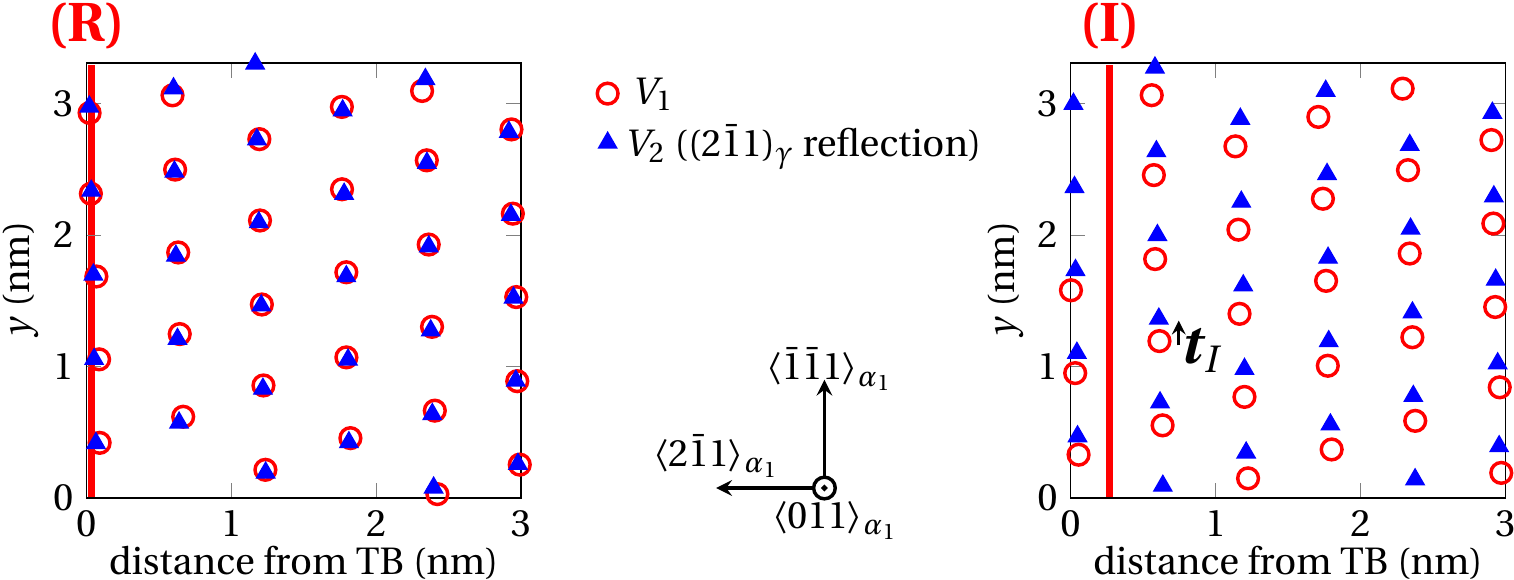}\label{fig:6b}}
\caption{Reflection (R) and isosceles (I) TB structure for the $(2\bar{1}1)_{\alpha_1}|\langle\bar{1}\bar{1}1\rangle_{\alpha_1}$ twinning mode, projected on the $(011)_{\alpha}$ plane. (I) TB obtained from (R) TB by imposing an additional translation of $\vect{t}_I=\langle\bar{1}\bar{1}1\rangle_{\alpha_1}$ at the interface. (a) Schematic representation. Dark blue: $V_2$ bcc atoms on the $(011)_{\alpha}$ plane. Red: $V_1$ bcc atoms in the $(011)_{\alpha}$ plane.  Light blue: $V_2$ bcc atoms in the plane above and beneath the $(011)_{\alpha}$ plane. Pink: $V_1$ bcc atoms in the plane above and beneath the $(011)_{\alpha}$ plane. (b) TB as simulated by QA at $t=1500$. Red circles: $V_1$ atoms in the $(011)_{\alpha}$ plane. Blue triangles: $V_2$ atoms  in the $(011)_{\alpha}$ plane after $(1\bar{2}1)_{\gamma}$ plane reflection. }
\label{fig:6} 
\end{figure}

The twinned domains $V_1$ and $V_2$ envisioned in figure  \ref{fig:2} are connected by twin boundaries (TB) lying parallel to the $V_1$/$V_2$ twinning plane  $K_1=(2\bar{1}1)_{\alpha_1}$. Two such TB are displayed in figure \ref{fig:5a} and \ref{fig:5b}. Their location  in the $(011)_{\alpha_1}$ plane is given by the row of gray atoms in OVITO's CNA (perturbed structure) which marks the transition from one bcc variant to another. The more accurate location of the TB along gray atoms can  be further determined, depending on the type of TB (red dashes in figures \ref{fig:5a} and \ref{fig:5b}).  In bcc systems, TB can be of two types: reflection (R) and isosceles (I) \cite{bilby1965theory} as sketched in figure \ref{fig:6a}. In the first case (figure \ref{fig:6a}, left), the twinned structure is obtained by reflection in the twinning plane $(2\bar{1}1)_{\alpha_1}$, and the TB is precisely spotted at the reflection plane which cuts the center of gray atoms. In the second case (figure \ref{fig:6a}, right), the atoms belonging to one variant are translated by a vector $\vect{t}_I=\frac{1}{12}\langle\bar{1}\bar{1}1\rangle_{\alpha_1}$ compared to the reflection TB, and the mirror symmetry is violated. In that case, the TB is located between the  $(2\bar{1}1)_{\alpha_1}$ twinning plane and the next  $(2\bar{1}1)_{\alpha_1}$ plane.

In this work, we mostly observed isosceles (I) twin boundaries (figure \ref{fig:5a} --top and bottom-- and figure \ref{fig:5b}). The isosceles nature of the top TB portion in figure \ref{fig:5a} is demonstrated in figure \ref{fig:6b} (right), where atoms belonging to  variant $V_2$  (blue triangles) are superimposed to atoms belonging to variant $V_1$  (red squares) in the  $(011)_{\alpha_1}$ plane, after $(2\bar{1}1)_{\alpha_1}$ plane reflection. As a result, atoms belonging to variants  $V_1$ and $V_2$ are shifted along the $\langle\bar{1}\bar{1}1\rangle_{\alpha_1}$ direction by the translation vector $\vect{t}_I$. Portions of reflection (R) twin boundaries were also identified (strip of TB in the center of figure \ref{fig:5a}). Using the same procedure as for the isosceles TB, figure \ref{fig:6b} evidences the mirror symmetry between the two variants close to this portion of TB, insofar as atoms belonging to $V_1$ and $V_2$ superimpose after $(2\bar{1}1)_{\alpha_1}$ plane reflection.

During propagation of TB, the transition from one type of TB to another --(I)$\to$(R) and (R)$\to$ (I)-- is accompanied by the local re-stacking of $(2\bar{1}1)_{\alpha_1}$ planes, which results in the shift of the interface along the direction $\langle 2\bar{1}1\rangle_{\alpha_1}$ perpendicular to the twinning plane. In the present work, the amplitude of the shift was observed to be half the distance between two successive $(2\bar{1}1)_{\alpha_1}$ planes. This is materialized by the formation of a step with height $h=h_{IR/RI}=a_2/(2\sqrt{6})$, as indicated by the mismatch between red dashed lines indicating the TB in figure \ref{fig:5a}. 

TB steps also bear a dislocation nature referred to as twinning dislocation \cite{vitek1970core,bristowe1976zonal,bristowe1977computer,christian1995deformation}. In figure \ref{fig:5a}, the two steps  correspond to two partial twin dislocations (IR and RI). Each partial is framed by its Burgers circuit (black dashed line in figures \ref{fig:5a} and \ref{fig:5b}) traced out around the dislocation in the FS/RH reference crystal convention  \cite{bristowe1977computer}. The calculated Burgers vector $\vect{b}$ is close to the theoretical partial IR twin dislocation: $\vect{b}\simeq\vect{b}_{\text{IR/RI}}=\frac{1}{12}\langle\bar{1}\bar{1}1\rangle_{\alpha_1}$. The correspondence  $\vect{b}_{\text{IR/RI}}=\vect{t}_I$ indicates that (IR) and (RI) partial twin dislocations carry the translation of atoms close to the TB, when switching from (I) to (R) and (R) to (I) types of TB. Moreover, in the present simulations, the vast majority of partial twin dislocations form (IR+RI) pairs connected by a strip of reflection (R) interface, and  bordered by two strips of isosceles (I) interface (figure \ref{fig:5a}). From the general theory of twin dislocations \cite{bristowe1977computer,christian1995deformation}, we surmise that the two partial dislocations having the same Burgers vector repel one another, up to a distance where the increase of (R) type TB energy balances the repulsion between partials.

Regarding the dynamics of twin variant growth, we could observe that the propagation of the TB exclusively proceeds by the glide of pairs (IR+RI) of partial twin dislocations (see  movies 1 and 2 of the Online Supplementary Material), which translates TB steps along the $\langle\bar{1}\bar{1}1\rangle_{\alpha_1}$ direction parallel to $\vect{b}_{\text{IR/RI}}$. The strong glissility of partial twin dislocations which roots the propagation of the TB, stems from their wide core (diffuse step) (see figure \ref{fig:5a}). As a matter of fact, it was shown in \cite{shi2016competing} that the double $\frac{1}{12}\langle\bar{1}\bar{1}1\rangle_{\alpha}$ dislocation glide was the prominent TB propagation in different metals, including bcc Iron. A more detailed description of this twin partials glide process can also be found there.

Close to the fcc/bcc interface, zonal dislocations (Z) characterized by the Burgers vector  $\vect{b}_{\text{Z}}=\frac{1}{6}\langle\bar{1}\bar{1}1\rangle_{\alpha_1}$, and a step height of two $(2\bar{1}1)_{\alpha_1}$ interplanar spacing could seldom be observed, as displayed in figure \ref{fig:5b}. However, zonal dislocations are usually associated to a different twinning mode in bcc systems \cite{rowlands1970application} than the present $(2\bar{1}1)_{\alpha_1}|\langle\bar{1}\bar{1}1\rangle_{\alpha_1}$ mode. The emergence of (Z) dislocations might thus be induced by the stress exerted by the near fcc/bcc interface. From the kinetic perspective, zonal dislocations are poorly glissile, and their contribution to TB propagation is thus negligible.

\subsection{fcc/bcc interface}

\paragraph{Surface relief of the fcc/bcc interface}
 
\begin{figure}[ht]
\centering
\includegraphics[width=14cm]{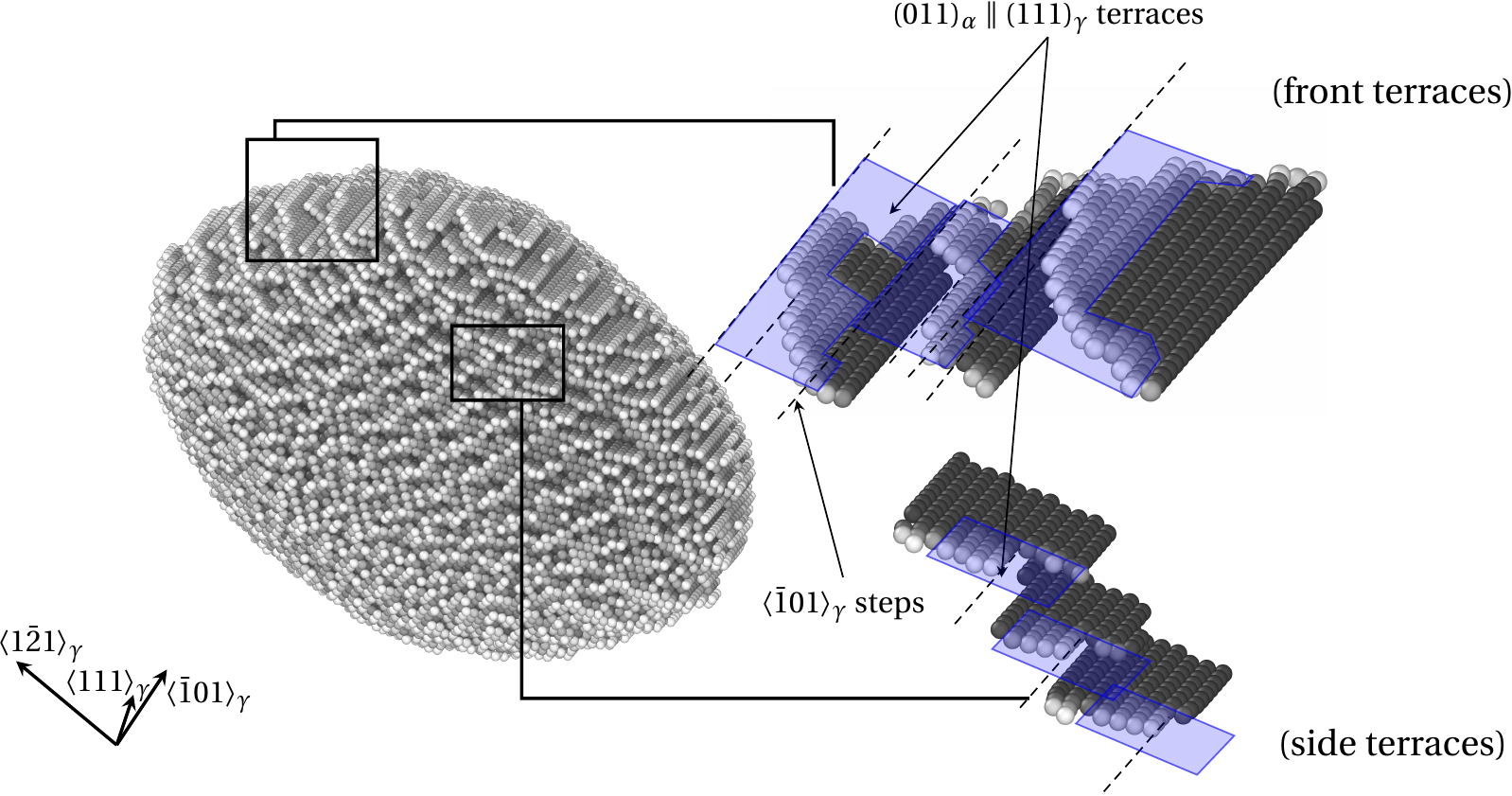}
\caption{Surface relief of the bcc inclusion (bcc+unknown+hcp atoms as provided by OVITO CNA) as extracted from QA simulations at $t=1500$, using OVITO's 'ambient occlusion' visual rendering. Top (top right) and side (bottom right) coherent units consisting in $(011)_{\alpha}\parallel(111)_{\gamma}$ terraces are spotted by transparent blue planes. Shades of gray reflect the distance of atoms from the fcc/bcc interface. }
\label{fig:7} 
\end{figure}

The bcc inclusion extracted from QA simulations at $t=1500$  is presented in figure \ref{fig:7}, using OVITO's visual rendering to magnify the surface relief. It was obtained by removing all atoms with fcc first neighbor environment, so that atoms labeled with bcc, hcp and perturbed structures were considered to belong to the inclusion in this figure.

The relief of the inclusion takes the form of a series of surface units consisting in $(111)_{\gamma}\parallel(011)_{\alpha}$ terraces, separated by $\langle \bar{1}01\rangle_{\gamma}$ steps of one or  several $(111)_{\gamma}$ interplanar spacings. Three $(011)_{\alpha}\parallel(111)_{\gamma}$ planes taken on the top of the inclusion (top terrace) where the average orientation of the interface is close to the $(011)_{\alpha}$ plane are enlarged on the top right of figure \ref{fig:7}. The corresponding terraces are flagged  by transparent blue planes. The same is done for three terraces spotted more on the side of the inclusion (side terrace, at the bottom right of figure \ref{fig:7}). Therefrom, it appears that $(111)_{\gamma}$ terraces can be found all over the surface of the inclusion. However, the length of these terraces is bigger when the average orientation of the interface is close to the $(011)_{\alpha}$ terrace plane (top and bottom of the inclusion), and smaller otherwise (side of the inclusion).

\begin{figure}[ht]
\centering
\subfigure[~~Atomic rhombi]{\includegraphics[width=8.5cm]{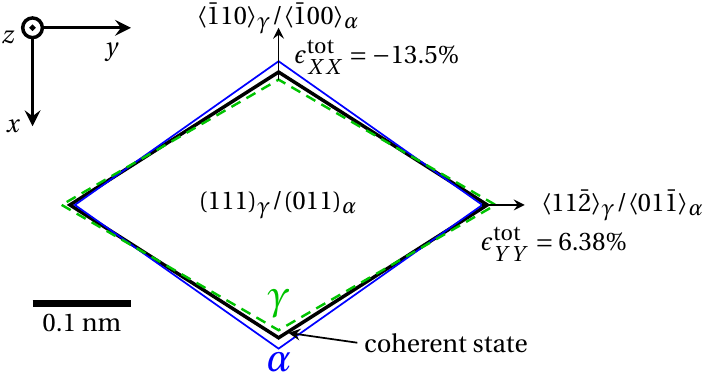}\label{fig:8a}}
\subfigure[~~Strain]{\includegraphics[width=7.5cm]{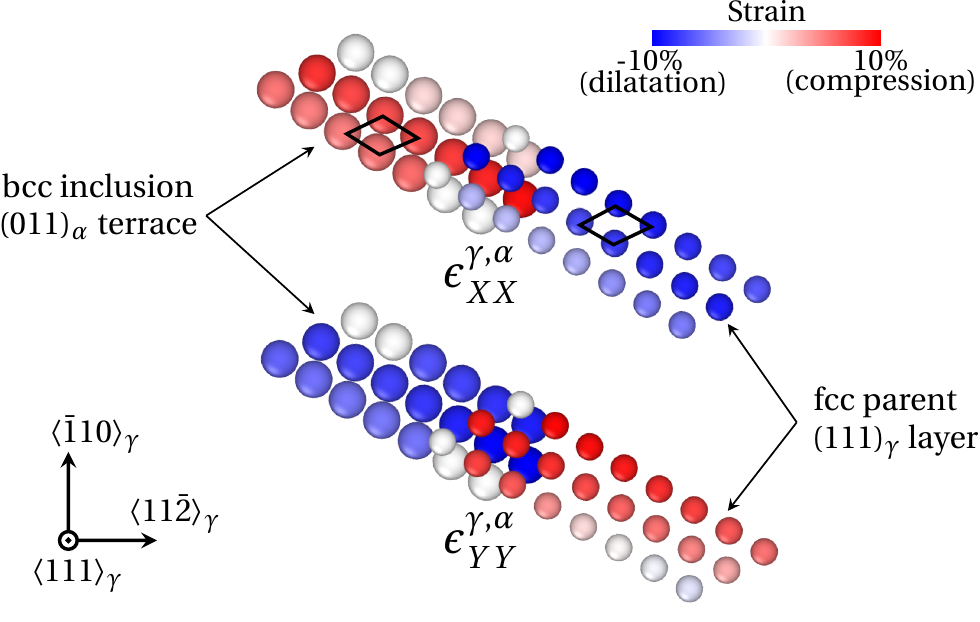}\label{fig:8b}}
\caption{Misfit strain at the coherent $(011)_{\alpha}\parallel (111)_{\gamma}$ inclusion terrace expressed in the terrace plane coordinate frame $(\langle\bar{1}10\rangle_{\gamma},\langle 11\bar{2}\rangle_{\gamma},\langle 111\rangle_{\gamma}$). (a) Scale drawing of the atomic rhombi in the terrace plane. Blue lines represent the unstrained bcc structure, green dashed lines the unstrained parent crystal, and bold black lines the coherent state. (b) QA calculation ($t=1500$)  at one $(011)_{\alpha}$ coherent terrace. Top: $\epsilon_{XX}^{\gamma,\alpha}$. Bottom: $\epsilon_{YY}^{\gamma,\alpha}$.}
\label{fig:8} 
\end{figure}

In virtue the TM \cite{pond2003comparison}, the fcc/bcc interface with KS OR should be fully coherent at $(011)_{\alpha}\parallel(111)_{\gamma}$ terraces. Under this hypothesis, the total misfit strain at the coherent $(011)_{\alpha}$ terrace can be calculated in the terrace plane coordinate frame $(\langle\bar{1}10\rangle_{\gamma},\langle 11\bar{2}\rangle_{\gamma},\langle 111\rangle_{\gamma}$) as:

\begin{equation}
S_t=
\begin{pmatrix}
\epsilon_{XX}^{\text{tot}}& 0& 0\\
0&\epsilon_{YY}^{\text{tot}}&0\\
0 & 0 & 0
\end{pmatrix},
\label{eqstrain}
\end{equation}
where both the deformation of the bcc and fcc phases are accounted for in the total strain. In the present work, the lattice parameter of the unstrained fcc and bcc structures are $a_1=0.36$ nm and $a_2=0.293$ nm respectively, so that in theory, $\epsilon_{XX}^{\text{tot}}\simeq 2(d_{XX}^{\alpha}-d_{XX}^{\gamma})/(d_{XX}^{\alpha}+d_{XX}^{\gamma})=-13.5$\% and $\epsilon_{YY}^{\text{tot}}\simeq 2(d_{YY}^{\alpha}-d_{YY}^{\gamma})/(d_{YY}^{\alpha}+d_{YY}^{\gamma})=6.38$\% in the  $(011)_{\alpha}\parallel (111)_{\gamma}$ terrace plane. Here, $d_{XX}^{\alpha}$, $d_{YY}^{\alpha}$, $d_{XX}^{\gamma}$ and $d_{YY}^{\gamma}$ are  the interatomic distances for the  fcc and bcc structures, in the  $\langle \bar{1}00\rangle_{\alpha}\parallel\langle \bar{1}10\rangle_{\gamma}$ and  $\langle 11\bar{2}\rangle_{\gamma}\parallel\langle01\bar{1}\rangle_{\alpha}$ directions respectively. A strong dilatation (compression) of the fcc (bcc) structure in the $\langle \bar{1}10\rangle_{\gamma}\parallel\langle \bar{1}00\rangle_{\alpha}$ (X) direction, and  compression (dilatation) of the fcc (bcc) structure in the $\langle 11\bar{2}\rangle_{\gamma}\parallel\langle01\bar{1}\rangle_{\alpha}$ (Y) direction should thus be observed at coherent terraces. This is  summarized in figure \ref{fig:8a} where the rhombi delineating the atomic sites on the $ (011)_{\alpha}\parallel (111)_{\gamma}$ planes for the unstrained fcc (green dashes) and bcc (blue line) structures and the coherent state (black line) are depicted.

To ascertain the coherency of $(011)_{\alpha}$ terraces in the present QA simulations, the deformation of the fcc and bcc structures with respect to the unstrained state were calculated at these terraces. Therefrom, the two principal components XX and YY of the fcc and bcc strains as expressed in the terrace referential plane are displayed in figure \ref{fig:8b} for $t=1500$, wherein big spheres correspond to  atoms with bcc environment and belonging to the $(011)_{\alpha}$ plane, and small spheres correspond to atoms with fcc environment and belonging to the interfacial $(011)_{\alpha}\parallel(111)_{\gamma}$ plane. Moreover, the colormap reflects the amplitude of the accommodation strain for each phase.  In the $XX$ direction, the bcc phase is strongly compressed while the fcc structure is dilated. In the $YY$ direction, it is the opposite situation (see figure \ref{fig:8a}). 


%

\paragraph{Defect structure of the fcc/bcc interface}

\begin{figure}[H]
\centering
\includegraphics[width=16cm]{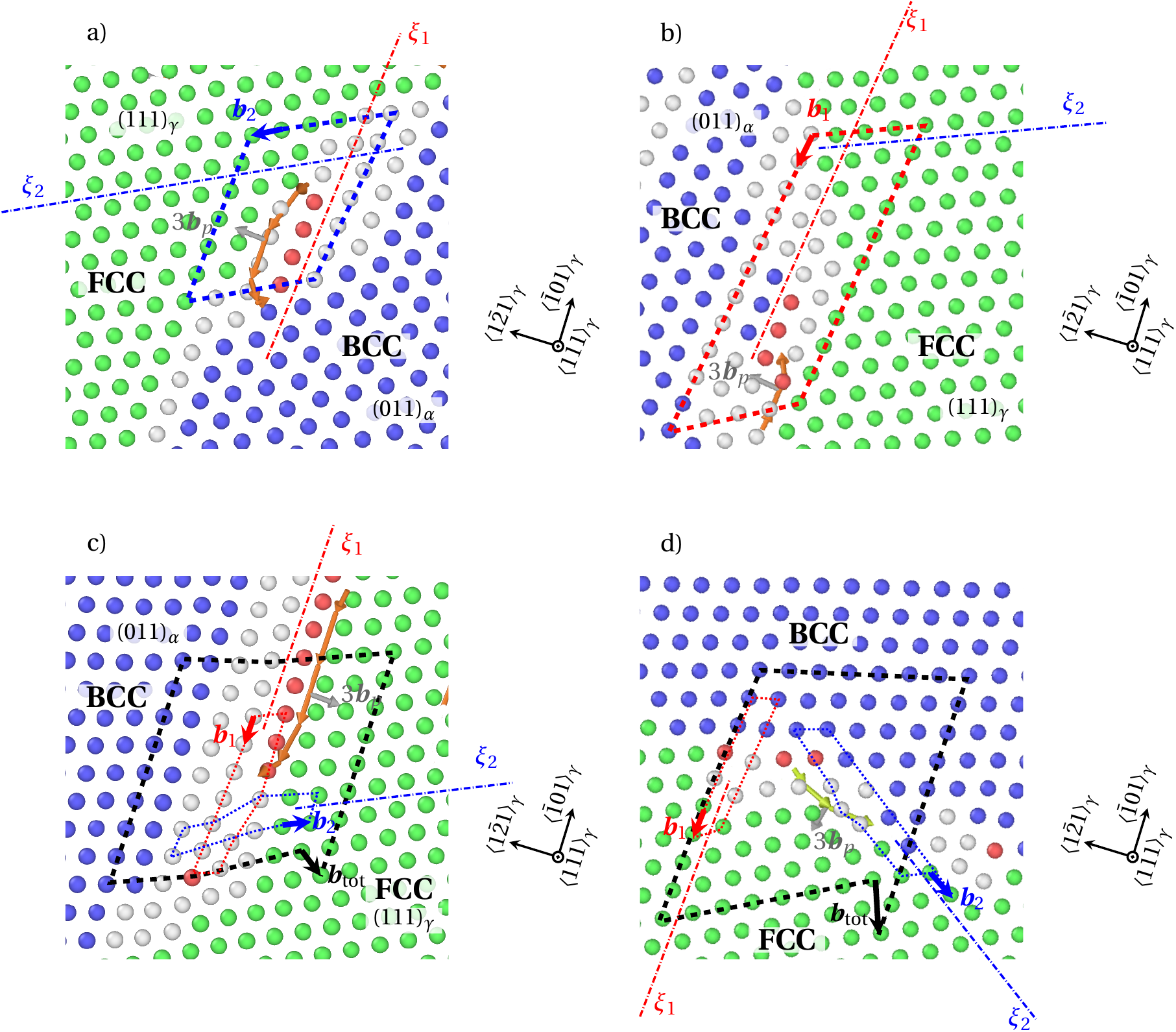}
\caption{Defect structure of the fcc/bcc interface of  $(011)_{\alpha}$ terraces, as extracted from QA simulations at $t=750$. Visual rendering uses OVITO's CNA and Diffraction analysis in the fcc structure.  1/ Along front edge ($\langle\bar{1}01\rangle_{\gamma}$ direction): (fcc) stacking fault (SF) marked by $\langle\bar{1}01\rangle_{\gamma}$ rows of hcp atoms, bordered by  Shockley partial dislocations  (orange line) with Burgers vector $\vect{b}_p=\pm \frac{1}{6} \langle 1\bar{2}1\rangle_{\gamma}$ (gray arrow), and accompanied by fcc screw dislocation with Burgers vector $\vect{b}_1=\pm\frac{1}{2}\langle\bar{1}01\rangle_{\gamma}$ (red arrow) and dislocation line $\xi_1=\langle\bar{1}01\rangle_{\gamma}$ (red dash-dotted line). 2/ Along side edge ($\langle 1\bar{1}0\rangle_{\gamma}/\langle 0\bar{1}1\rangle_{\gamma}$ directions): (fcc) SF  + Shockley partials (yellow tube) with $\vect{b}_p=\pm \frac{1}{6} \langle \bar{1}12\rangle_{\gamma}$ (gray arrow), siding fcc screw dislocation with $\vect{b}_2=\pm\frac{1}{2}\langle 1\bar{1}0\rangle_{\gamma}$ and $\pm\frac{1}{2}\langle 0\bar{1}1\rangle_{\gamma}$ (blue arrow) and $\xi_2=\langle\bar{1}01\rangle_{\gamma}/\langle 0\bar{1}1\rangle_{\gamma}$ (blue dash-dotted line). a) Burgers circuit for $\vect{b}_2$ dislocations only (blue dashed line). b) Burgers circuit for $\vect{b}_1$ dislocations only (red dashed line). c) and d) Burgers circuit for resulting dislocation $\vect{b}_{\text{tot}}$ (black dashed line), and Burgers circuit for $\vect{b}_2$ and $\vect{b}_1$ contributions (blue and red dotted line).}
\label{fig:10} 
\end{figure}

\begin{figure}[ht]
\centering
\includegraphics[width=10cm]{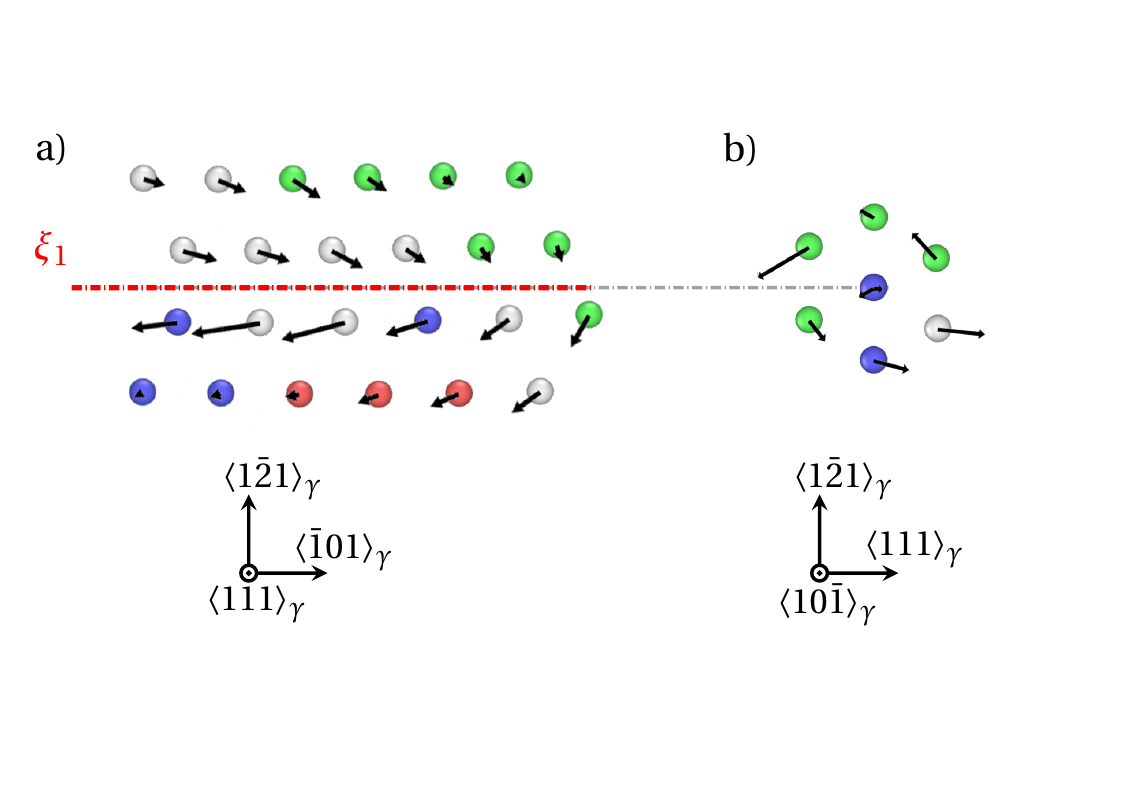}
\vspace{-1.5cm}
\caption{$\vect{b}_1$ screw dislocation at the fcc/bcc interface step of $(011)_{\alpha}$ terraces, as extracted from QA simulations at $t=800$. Visual rendering uses OVITO's CNA. Black arrows represent the displacement field (amplitude $\times 50$) of atoms, between close times $t=800$ and $t=802.5$. (a) $\xi_1$ dislocation line (red dashed line) in the $(011)_{\alpha}\parallel(111)_{\gamma}$ plane. b) $(\bar{1}01)_{\gamma}$ plane slice of the step, aligned with (a). }
\label{fig:11} 
\end{figure}

To understand the mechanism rooting the propagation of the fcc/bcc interface,  the defect structure at this interface in the $(011)_{\alpha}\parallel (111)_{\gamma}$ plane is prospected in figures \ref{fig:10} and \ref{fig:11}. Therein, gray atoms  indicate the interphase structure connecting the fcc and the bcc phases, and green and blue spheres refer to atoms in the fcc and bcc structure, respectively. $(011)_{\alpha}$ terraces are schematically delineated by terrace edges aligned with the $\langle \bar{1}01\rangle_{\gamma}$ direction (front edge at the top right in figure \ref{fig:7}), and  the $\langle 1\bar{1}0\rangle_{\gamma}/\langle 0\bar{1}1\rangle_{\gamma}$ directions (side edge at the bottom right in figure \ref{fig:7}).

At front edges ($\langle\bar{1}01\rangle_{\gamma}$ direction), $\langle \bar{1}01\rangle_{\gamma}$ rows of atoms belonging to the hcp phase (in red) indicate the presence of $\langle \bar{1}01\rangle_{\gamma}$ stacking faults \cite{maresca2017austenite} produced by Shockley partial edge dislocations (orange tube) with Burgers vector $\vect{b}_p=\pm\frac{1}{6} \langle 1\bar{2}1\rangle_{\gamma}$ (gray arrow). The $\langle \bar{1}01\rangle_{\gamma}$ SF+$\frac{1}{6} \langle 1\bar{2}1\rangle_{\gamma}$ partial set is accompanied by a fcc screw dislocation with the Burgers vector $\vect{b}_1=\frac{1}{2}\langle \bar{1}01\rangle_{\gamma}$ (red arrow) and the dislocation line $\xi_1=\langle\bar{1}01\rangle_{\gamma}$ (red dashdotted line).  One $\vect{b}_1$ dislocation  is singled out in figure \ref{fig:10} b), where it is circumscribed by its Burgers circuit (red dashed line).

A similar structure is found at side edges ($\langle 1\bar{1}0\rangle_{\gamma}/\langle 0\bar{1}1\rangle_{\gamma}$ directions). Therein, Shockley partial edge dislocations (yellow tube) with Burgers vector $\vect{b}_p=\pm \frac{1}{6} \langle \bar{1}12\rangle_{\gamma}$ produce $\langle 1\bar{1}0\rangle_{\gamma}/\langle 0\bar{1}1\rangle_{\gamma}$ SF. This set of defects is again tied to a second class of fcc screw dislocations having the Burgers vector $\vect{b}_2=\pm\frac{1}{2}\langle 1\bar{1}0\rangle_{\gamma}/\pm\frac{1}{2}\langle 0\bar{1}1\rangle_{\gamma}$ (blue arrow) and the dislocation line $\xi_2=\langle\bar{1}01\rangle_{\gamma}/\langle 0\bar{1}1\rangle_{\gamma}$ (blue dashdotted line). One $\vect{b}_2$ dislocation  is isolated in figure \ref{fig:10} a), where it is framed by its Burgers circuit (blue dashed line).

The net Burgers vector $\vect{b}_{\text{tot}}$ (black arrow) of $\vect{b}_1$ and $\vect{b}_2$ dislocations is defined in figure \ref{fig:10} c) and \ref{fig:10} d) by the closure failure of the Burgers circuit (black dashed line) which encompasses both  $\vect{b}_1$ and $\vect{b}_2$ dislocations as circumscribed by their own Burgers circuits (blue and red dotted lines). The presence of similar screw dislocations at the fcc/bcc interface in Fe-0.6C-2Si-1Mn  and Fe-20Ni-5.5Mn steels was reported in  \cite{moritani2002comparison}.

\begin{figure}[ht]
\centering
\includegraphics[width=10cm]{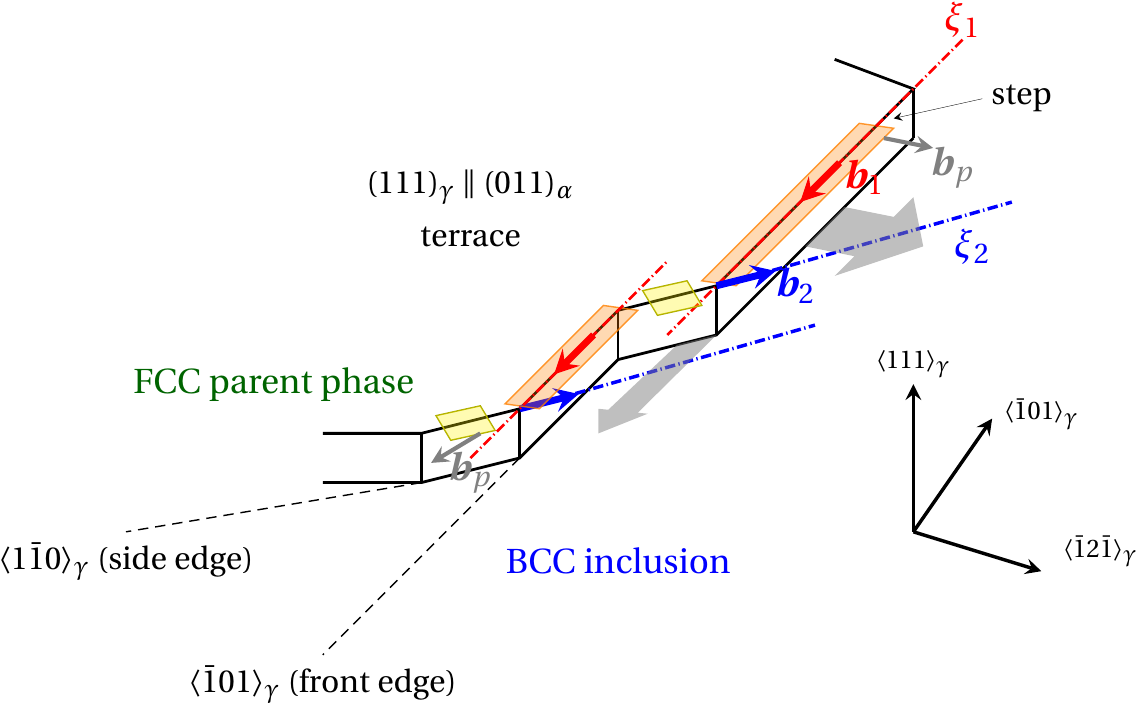}
\caption{Schematic representation of fcc/bcc interface structure and propagation mechanism in the case of an ellipsoidal bcc inclusion. One $(011)_{\alpha}$ terrace consisting in $\langle \bar{1}01\rangle_{\gamma}$ and $\langle 1\bar{1}0\rangle_{\gamma}/\langle 0\bar{1}1\rangle_{\gamma}$ step edges and corners is depicted. The fcc$\to$bcc interface is located at fcc SF (along $\langle \bar{1}01\rangle_{\gamma}$ in red, and $\langle 1\bar{1}0\rangle_{\gamma}/\langle 0\bar{1}1\rangle_{\gamma}$ in yellow) produced by fcc Shockley partial edge dislocations with Burgers vector $\vect{b}_p$. The terrace step propagates outward (gray arrows), by means of the glide of $\vect{b}_1$ (red arrow) screw dislocation along $\langle \bar{1}01\rangle_{\gamma}$ and $\vect{b}_2$ (blue arrow) screw dislocation along $\langle 1\bar{1}0\rangle_{\gamma}/\langle 0\bar{1}1\rangle_{\gamma}$.}
\label{fig:14} 
\end{figure}

Based on these observations, the structure and propagation mechanism of the fcc/bcc interface is in the case of an ellipsoidal bcc inclusion is schematized in figure  \ref{fig:14}, using the same color coding for dislocations as in figure \ref{fig:10}. At both front and side edges, the fcc$\to$bcc phase transition follows the fcc$\to$hcp$\to$bcc transformation path. It is triggered by $\vect{b}_p$ Shockley partial dislocations that produce SF along the edge, where the fcc/bcc interface can be spotted. This path can be divided into two steps: first, a Shockley partial dislocation with Burgers vector $\frac{1}{6} \langle 1\bar{2}1\rangle_{\gamma}$ produces a SF which transforms the ABCABC stacking sequence of close-packed $(111)_{\gamma}$ planes of the fcc structure into the ABAB stacking sequence in the hcp phase. Then, the homogeneous deformation of the hcp structure produces the final transformation from $(111)_{\gamma}$ fcc to $(011)_{\alpha}$ bcc planes. This homogeneous deformation is carried by the glide of $\vect{b}_1$ and $\vect{b}_2$ fcc screw dislocations along there respective dislocation line. In details, the glide generates a shear displacement of atomic positions along the edge in the $(011)_{\alpha}\parallel(111)_{\gamma}$ plane (figure \ref{fig:11} a)), while the rotation of atoms around the dislocation line (figure \ref{fig:11} b)) aligns the fcc perturbed structure on the bcc structure, in the vicinity of the step.  The present mechanism provides an athermal/glissile propagation mode for a curved fcc/bcc interface \cite{maresca2017austenite}.

%


\begin{figure}[ht]
\centering
\includegraphics[width=8cm]{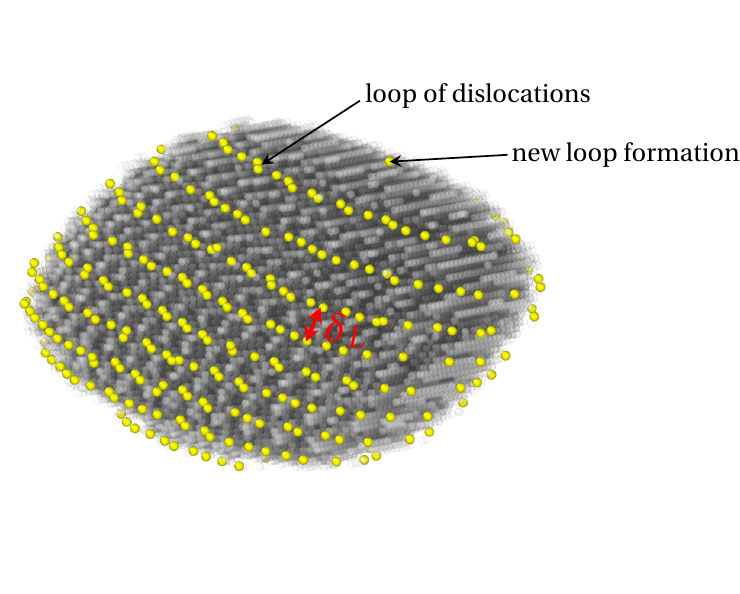}
\caption{a) 3D distribution of $\vect{b}_2$ dislocation cores at the interface of the inclusion (transparent gray atoms with ambient occlusion) as obtained by QA simulations at $t=750$. b) Expansion of one loop of dislocations  in the $(263)_{\gamma}$ plane.}
\label{fig:12} 
\end{figure}


Furthermore, $\vect{b}_2$ dislocations contribute to accommodate the misfit between the fcc structure in the $\langle 01\bar{1}\rangle_{\gamma}$ direction and the bcc structure in the $\langle 1\bar{1}1\rangle_{\alpha}$ (quasi-) parallel direction, for a curved fcc/bcc interface. One argument for that can be found in the three dimensional distribution of $\vect{b}_2$ dislocation cores (yellow spheres) on the surface of the ellipsoidal inclusion in figure \ref{fig:12} a).  These are organized in a lines of dislocation cores organized in loops which encircle the bcc inclusion, and lye in periodically distributed parallel planes. This type of defects was previously envisioned using a mixed MD-MC approach  at the semi-coherent interface of bcc-Cr precipitates in a fcc-Cu matrix, where it was referred to as "dislocation loops" \cite{dai2020coherent}. This periodicity can be related to the elastic accommodation of a twinned bcc particle with the fcc matrix. As was discussed in \cite{khachaturyan1991adaptive}, the period $\lambda$ of the twinned interface is the ratio between the conventional surface energy of the twin boundaries between two orientation variants, and the elastic energy related to semicoherent interface:

\begin{equation}
\label{period}
   \lambda=\sqrt{\frac{\gamma_{\text{TB}}}{\mu\epsilon_0}D},
\end{equation}
where $D$ is the width of the plate, $\gamma_{\text{TB}}$ is the twin surface energy, and $\mu$ is the shear modulus. In the present work, the periodic structure in figure \ref{fig:12} with period $\delta L=1.36$ nm can be also related to the ratio between the energy of staking faults and the elastic energy. Herein, the increase of the elastic energy or the decrease of the staking faults energy will result in the increase of $\delta L$. 


\subsection{Comparison with experimental results}

To compare our predictions with a real system, the Cu-Fe alloys seems appropriate (see section \ref{QAmodel}). In this immiscible system, the nanosized Fe-rich particles precipitate in the Cu-rich matrix during the cooling stage of the casting process. The crystallographic features of the Fe-rich nanoparticle in the Cu-2.0Fe-0.5Co (wt. \%) alloy was investigated via transmission electron microscopy (TEM). Details about experimental conditions are provided in the appendix \ref{exp}. In figure  \ref{fig:13} (a-c), a spherical Fe-rich nanoparticle is characterized by high resolution TEM (HRTEM) and Fast Fourier Transform (FFT) pattern along the $\langle 111\rangle_{\alpha}$ zone axis (ZA), indicating a bcc structure of Fe-rich nanoparticle and its KS OR with  the fcc Cu matrix. Along the $\langle \bar{1}13\rangle_{\alpha}$ ZA of the same Fe-rich nanoparticle, a bcc twinned domains within the nanoparticle is dsplayed. It is accompanied by HRTEM exhibition and FFT pattern indexation (figure \ref{fig:13} (d-f). The amplified image of the twin boundary zone in figure \ref{fig:13} f) presents partial twin dislocation couple (IR+RI) corresponding to the numerical atomic structure displayed in figure \ref{fig:5} a). In addition, the FFT pattern of one Fe-rich nanoparticle along the $\langle 011\rangle_{\alpha}$ ZA indicates that twinning domains are formed with the KS OR variants V1 and V2 (figure \ref{fig:13} (g-i). This is similar to the present QA simulations (figures \ref{fig:2} and \ref{fig:4}). The theoretical misorientation angle between the KS variants V1 and V2 observed experimentally (70.62$^{\circ}$, measured using e-Ruler 1.0) in figure \ref{fig:13} i) is also very close to the numerical value (70.5$^{\circ}$). The experimental thickness of twin domains is avg. 8.0 nm (ranging from 2.2 to 23.6 nm) vs. 15 nm in simulations. This disparity stems from the fact that the simulation box in our simulations is smaller than the size of Fe-rich particle observed experimentally. However, it is shown in this study that this type of structures can be also observed in smaller particles using scaling parameters. It is also noteworthy that a bcc envelope emerges between one spheroidal bcc (Iron) precipitate and the fcc (Copper) matrix in figure \ref{fig:13} a) and d). We suggest that it might correspond to the transient envelope obtained in simulations (figure \ref{fig:2}) between the bcc inclusion and the fcc parent phase.

\begin{figure}[ht]
\centering
\includegraphics[height=14cm]{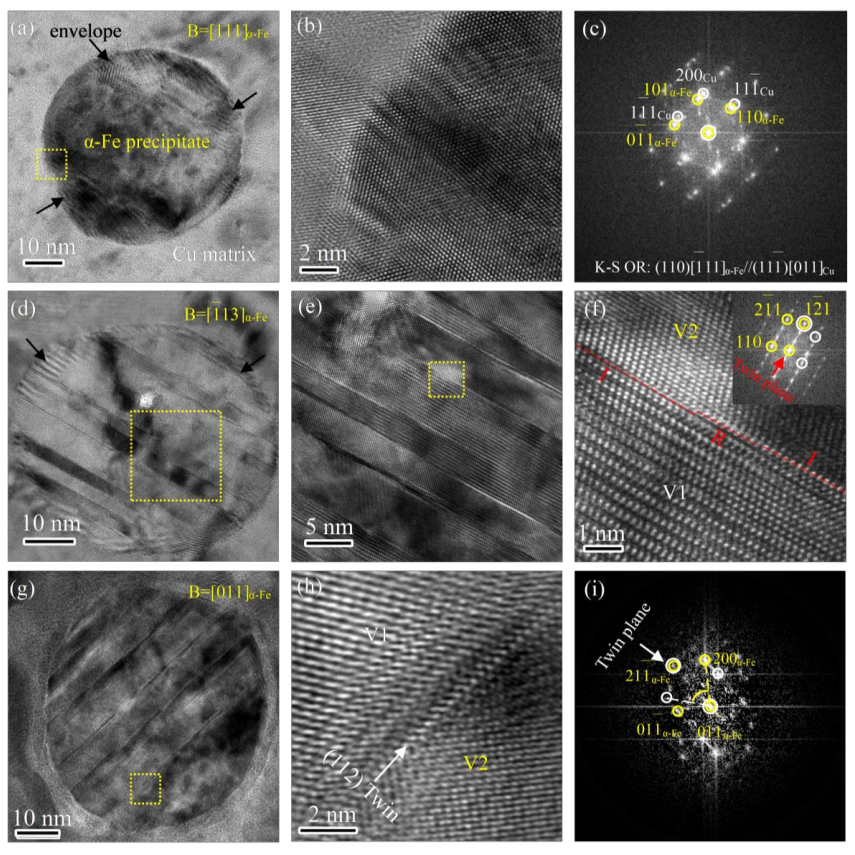}
\caption{HRTEM analysis of one spherical Iron-rich nanoparticle along different directions. a) HRTEM micrograph along $\langle 111\rangle_{\alpha}$ zone axis. b) Enlarged figure of dashed squire zone in a). c) the corresponding FFT pattern of b), presenting the KS OR between the Iron-rich nanoparticle and the copper matrix. d) HRTEM micrograph along $\langle \bar{1}13\rangle_{\alpha}$ zone axis, showing twinning domains. e) Enlarged figure of dashed squire zone in d). f) Zoom on the one of the twin boundaries marked by dashed squire zone in e) with partial twin dislocation couple (IR+RI), the inset showing its corresponding FFT pattern. Envelopes of the iron-rich nanoparticle were marked by black arrows in (a, d). g) HRTEM micrograph along $\langle 011\rangle_{\alpha}$ zone axis. h) Amplified figure of dashed squire zone in g). i) Corresponding FFT pattern of h), presenting the KS OR V2 (yellow), and the KS OR V1
(White).}
\label{fig:13} 
\end{figure} 	

As a conclusion, while a fully quantitative comparison between our experimental and modeling results is still beyond reach, it appears that the present numerical model finely reproduces the main characteristics of the fcc$\to$bcc transformation in binary system. 

\section*{Conclusion}

In this work, the QA was further developed to simulate the fcc$\to$bcc transformation in a model binary system. At microstructural scale, the bcc inclusion was found to grow from a bcc nuclei with KS OR, in the shape of a slightly flattened ellipsoid. An effective habit plane $(575)_{\gamma}$ which minimizes the bulk strain energy of the inclusion was determined as the plane normal to the flattened direction. Besides, the growth of the inclusion was accompanied by the spontaneous appearance of the second KS variant with twin related OR. The simulated diffraction pattern was used to identify the nature of this second variant and to determine the twinning plane and twinning direction. It was shown that the corresponding twinning mode is $(2\bar{1}1)_{\alpha}|\langle\bar{1}\bar{1}1\rangle_{\alpha}$. At atomic level, twin boundaries were first found to be mostly of isosceles nature, albeit hosting strips of reflection boundaries between pairs of partial twin dislocations gliding along direction $\langle \bar{1}\bar{1}1\rangle_{\alpha}$.  We concluded that the prominent mechanism of twin growth was the glide of dissociated $\langle\bar{1}\bar{1}1\rangle_{\alpha}$ twin dislocations along twin boundaries. Another feature  predicted from our simulation is that the fcc/bcc interface of the bcc inclusion within the precipitate was found to consist in $(111)_{\gamma}$ coherent terraces delineated by steps. The defect structure of terrace steps was found to consist in pairs of $\frac{1}{2}\langle\bar{1}0\bar{1}\rangle_{\gamma}$ and $\frac{1}{2}\langle 1\bar{1}0\rangle_{\gamma}$ fcc screw dislocations. Therefrom, we could posit that the fcc/bcc interface athermal propagation was mainly carried by the glide of $\frac{1}{2}\langle\bar{1}0\bar{1}\rangle_{\gamma}$ screw dislocations along $\langle\bar{1}01\rangle_{\gamma}$ step edges, striding the fcc$\to$hcp$\to$bcc transformation path. Meanwhile, $\frac{1}{2}\langle 1\bar{1}0\rangle_{\gamma}$ dislocations were assumed to accommodate the misfit between the fcc and bcc structures in the $\langle 01\bar{1}\rangle_{\gamma}$ and $\langle 1\bar{1}1\rangle_{\alpha}$ directions respectively. Finally, a first comparison with HRTEM observations of twinned Iron rich precipitates in a cast Fe-Cu alloy proved qualitatively consistent with the present simulations.

As a conclusion, the present work demonstrates the potential of the QA to address the challenging issue of martensitic transformation, which is notoriously difficult to prospect experimentally. Herein, the fcc$\to$bcc transformation could be simulated in a large-scale system with complex structure patterns, while the atomic processes rooting the transformation could be identified. Also, an encouraging qualitative agreement between QA simulations and our experimental observations could be obtained. It is worth noting that the present formulation of the QA is fairly general, and paves the way for further studies, among which the propagation of a flat fcc/bcc interface and its interaction with solute atoms. 

\section*{DATA AVAILABILITY}

All datasets generated in the current study are available from the corresponding authors upon reasonable request.

\bibliographystyle{unsrt} 

\begin{thebibliography}{60}

\bibitem{olson1976general}
GB~Olson and Morris Cohen.
\newblock A general mechanism of martensitic nucleation: Part ii. fcc $\to$ bcc
  and other martensitic transformations.
\newblock {\em Metallurgical transactions A}, 7(12):1905--1914, 1976.

\bibitem{moritani2002comparison}
T~Moritani, N~Miyajima, T~Furuhara, and T~Maki.
\newblock Comparison of interphase boundary structure between bainite and
  martensite in steel.
\newblock {\em Scripta materialia}, 47(3):193--199, 2002.

\bibitem{sandvik1983characteristicsII}
BPJ Sandvik and CM~Wayman.
\newblock Characteristics of lath martensite: Part ii. the martensite-austenite
  interface.
\newblock {\em Metallurgical Transactions A}, 14(4):823--834, 1983.

\bibitem{sandvik1983characteristicsIII}
BPJ Sandvik and CM~Wayman.
\newblock Characteristics of lath martensite: Part iii. some theoretical
  considerations.
\newblock {\em Metallurgical Transactions A}, 14(4):835--844, 1983.

\bibitem{kelly1990orientation}
PM~Kelly, A~Jostsons, and RG~Blake.
\newblock The orientation relationship between lath martensite and austenite in
  low carbon, low alloy steels.
\newblock {\em Acta Metallurgica et Materialia}, 38(6):1075--1081, 1990.

\bibitem{qi2014microstructure}
Liang Qi, AG~Khachaturyan, and JW~Morris~Jr.
\newblock The microstructure of dislocated martensitic steel: Theory.
\newblock {\em Acta materialia}, 76:23--39, 2014.

\bibitem{ma2007parent}
X~Ma and RC~Pond.
\newblock Parent--martensite interface structure in ferrous systems.
\newblock {\em Journal of nuclear materials}, 361(2-3):313--321, 2007.

\bibitem{bos2006molecular}
C~Bos, J~Sietsma, and BJ~Thijsse.
\newblock Molecular dynamics simulation of interface dynamics during the
  fcc-bcc transformation of a martensitic nature.
\newblock {\em Physical Review B}, 73(10):104117, 2006.

\bibitem{ou2016molecular}
Xiaoqin Ou, Jilt Sietsma, and Maria~Jesus Santofimia.
\newblock Molecular dynamics simulations of the mechanisms controlling the
  propagation of bcc/fcc semi-coherent interfaces in iron.
\newblock {\em Modelling and Simulation in Materials Science and Engineering},
  24(5):055019, 2016.

\bibitem{wechsler1953theory}
Monroe~S Wechsler, DS~Lieberman, and TA~Read.
\newblock On the theory of the formation of martensite.
\newblock {\em Trans. Aime}, 197(11):1503--1515, 1953.

\bibitem{bowles1954crystallography}
JS~Bowles and JK~Mackenzie.
\newblock The crystallography of martensite transformations i.
\newblock {\em Acta metallurgica}, 2(1):129--137, 1954.

\bibitem{a1964introduction}
CM~{\'a}~Wayman.
\newblock Introduction to the crystallography of martensite transformations,
  1964.

\bibitem{khachaturyan2013theory}
Armen~G Khachaturyan.
\newblock {\em Theory of structural transformations in solids}.
\newblock Courier Corporation, 2013.

\bibitem{pond2003comparison}
RC~Pond, S~Celotto, and JP~Hirth.
\newblock A comparison of the phenomenological theory of martensitic
  transformations with a model based on interfacial defects.
\newblock {\em Acta materialia}, 51(18):5385--5398, 2003.

\bibitem{hirth2011compatibility}
JP~Hirth and RC~Pond.
\newblock Compatibility and accommodation in displacive phase transformations.
\newblock {\em Progress in Materials Science}, 56(6):586--636, 2011.

\bibitem{hirth1994dislocations}
JP~Hirth.
\newblock Dislocations, steps and disconnections at interfaces.
\newblock {\em Journal of Physics and Chemistry of Solids}, 55(10):985--989,
  1994.

\bibitem{hirth2016disconnections}
JP~Hirth, J~Wang, and CN~Tom{\'e}.
\newblock Disconnections and other defects associated with twin interfaces.
\newblock {\em Progress in Materials Science}, 83:417--471, 2016.

\bibitem{maresca2017austenite}
F~Maresca and WA~Curtin.
\newblock The austenite/lath martensite interface in steels: Structure,
  athermal motion, and in-situ transformation strain revealed by simulation and
  theory.
\newblock {\em Acta Materialia}, 134:302--323, 2017.

\bibitem{song2013atomistic}
H~Song and JJ~Hoyt.
\newblock An atomistic simulation study of the migration of an
  austenite--ferrite interface in pure {Fe}.
\newblock {\em Acta materialia}, 61(4):1189--1196, 2013.

\bibitem{castan1989kinetics}
Teresa Cast{\'a}n and Per-Anker Lindg{\aa}rd.
\newblock Kinetics of domain growth, theory, and monte carlo simulations: A
  two-dimensional martensitic phase transition model system.
\newblock {\em Physical Review B}, 40(7):5069, 1989.

\bibitem{chen2015coupled}
Ying Chen and Christopher~A Schuh.
\newblock A coupled kinetic monte carlo--finite element mesoscale model for
  thermoelastic martensitic phase transformations in shape memory alloys.
\newblock {\em Acta Materialia}, 83:431--447, 2015.

\bibitem{johnson1989analytic}
RA~Johnson and DJ~Oh.
\newblock Analytic embedded atom method model for bcc metals.
\newblock {\em Journal of Materials Research}, 4(5):1195--1201, 1989.

\bibitem{ackland1997computer}
GJ~Ackland, DJ~Bacon, AF~Calder, and T~Harry.
\newblock Computer simulation of point defect properties in dilute {Fe-Cu}
  alloy using a many-body interatomic potential.
\newblock {\em Philosophical Magazine A}, 75(3):713--732, 1997.

\bibitem{meyer1998martensite}
R~Meyer and P~Entel.
\newblock Martensite-austenite transition and phonon dispersion curves of
  {Fe$_{1- x}$ Ni$_x$} studied by molecular-dynamics simulations.
\newblock {\em Physical Review B}, 57(9):5140, 1998.

\bibitem{chen2002phase}
Long-Qing Chen.
\newblock Phase-field models for microstructure evolution.
\newblock {\em Annual review of materials research}, 32(1):113--140, 2002.

\bibitem{zhang2007phase}
W~Zhang, YM~Jin, and AG~Khachaturyan.
\newblock Phase field microelasticity modeling of heterogeneous nucleation and
  growth in martensitic alloys.
\newblock {\em Acta Materialia}, 55(2):565--574, 2007.

\bibitem{mamivand2013review}
Mahmood Mamivand, Mohsen~Asle Zaeem, and Haitham El~Kadiri.
\newblock A review on phase field modeling of martensitic phase transformation.
\newblock {\em Computational Materials Science}, 77:304--311, 2013.

\bibitem{jin2006atomic}
Yongmei~M Jin and Armen~G Khachaturyan.
\newblock Atomic density function theory and modeling of microstructure
  evolution at the atomic scale.
\newblock {\em Journal of applied physics}, 100(1):013519, 2006.

\bibitem{certain2011atomic}
Marilyne Certain, Helena Zapolsky, and Armen~G Khachaturyan.
\newblock Atomic density function simulations of crystal growth kinetics of fcc
  crystal and bcc-fcc transition.
\newblock In {\em Solid State Phenomena}, volume 172, pages 1234--1239. Trans
  Tech Publ, 2011.

\bibitem{demange2018generalization}
G~Demange, M~Chamaillard, H~Zapolsky, M~Lavrskyi, A~Vaugeois, L~Luneville,
  D~Simeone, and Renaud Patte.
\newblock Generalization of the fourier-spectral eyre scheme for the
  phase-field equations: Application to self-assembly dynamics in materials.
\newblock {\em Computational Materials Science}, 144:11--22, 2018.

\bibitem{lavrskyi2016quasiparticle}
Mykola Lavrskyi, Helena Zapolsky, and Armen~G Khachaturyan.
\newblock Quasiparticle approach to diffusional atomic scale self-assembly of
  complex structures: from disorder to complex crystals and double-helix
  polymers.
\newblock {\em Npj Computational Materials}, 2(1):1--9, 2016.

\bibitem{elder2002modeling}
KR~Elder, Mark Katakowski, Mikko Haataja, and Martin Grant.
\newblock Modeling elasticity in crystal growth.
\newblock {\em Physical review letters}, 88(24):245701, 2002.

\bibitem{elder2004modeling}
KR~Elder and Martin Grant.
\newblock Modeling elastic and plastic deformations in nonequilibrium
  processing using phase field crystals.
\newblock {\em Physical Review E}, 70(5):051605, 2004.

\bibitem{kapikranian2014atomic}
O~Kapikranian, H~Zapolsky, Ch~Domain, Renaud Patte, Cristelle Pareige,
  B~Radiguet, and Philippe Pareige.
\newblock Atomic structure of grain boundaries in iron modeled using the atomic
  density function.
\newblock {\em Physical Review B}, 89(1):014111, 2014.

\bibitem{kapikranian2015point}
O~Kapikranian, H~Zapolsky, R~Patte, C~Pareige, B~Radiguet, and Philippe
  Pareige.
\newblock Point defect absorption by grain boundaries in $\alpha$-iron by
  atomic density function modeling.
\newblock {\em Physical Review B}, 92(22):224106, 2015.

\bibitem{mavrikakis2019multi}
N~Mavrikakis, C~Detlefs, PK~Cook, M~Kutsal, APC Campos, M~Gauvin, PR~Calvillo,
  W~Saikaly, R~Hubert, Henning~Friis Poulsen, et~al.
\newblock A multi-scale study of the interaction of sn solutes with
  dislocations during static recovery in $\alpha$-fe.
\newblock {\em Acta Materialia}, 174:92--104, 2019.

\bibitem{chen_new}
KX~Chen, Pavel~A Korzhavyi, G~Demange, H~Zapolsky, Renaud Patte, Julien Boisse,
  and ZD~Wang.
\newblock Morphological instability of iron-rich precipitates in cufeco alloys.
\newblock {\em To be published}, 2020.

\bibitem{lavrskyi2017modelisation}
Mykola Lavrskyi.
\newblock {\em Mod{\'e}lisation en fonctionnelle de la densit{\'e} atomique des
  transformations de phases dans le syst{\`e}me Fe-C {\`a} basse
  temp{\'e}rature}.
\newblock PhD thesis, Normandie Universit{\'e}, 2017.

\bibitem{vaugeois2017modelisation}
Antoine Vaugeois.
\newblock {\em Mod{\'e}lisation de l'influence de la structure des joints de
  grains sur les ph{\'e}nom{\`e}nes de s{\'e}gr{\'e}gation.}
\newblock PhD thesis, 2017.

\bibitem{zarestky1987lattice}
J~Zarestky and C~Stassis.
\newblock Lattice dynamics of $\gamma$-fe.
\newblock {\em Physical Review B}, 35(9):4500, 1987.

\bibitem{lide2004crc}
David~R Lide.
\newblock {\em CRC handbook of chemistry and physics}, volume~85.
\newblock CRC press, 2004.

\bibitem{fratons2atoms}
A.~Goryaeva.
\newblock fratons2atoms.
\newblock \url{https://github.com/agoryaeva/fratons2atoms}, 2021.

\bibitem{eshelby1957determination}
John~Douglas Eshelby.
\newblock The determination of the elastic field of an ellipsoidal inclusion,
  and related problems.
\newblock {\em Proceedings of the royal society of London. Series A.
  Mathematical and physical sciences}, 241(1226):376--396, 1957.

\bibitem{fitzgibbon1996m}
F~Fitzgibbon and Pilu AW.
\newblock In {\em B “Direct least squares fitting of ellipses,” In Proc. of
  the 13th International Conference on Pattern Recognition, Vienna}, pages
  253--257, 1996.

\bibitem{krauss2015steels}
George Krauss.
\newblock {\em Steels: processing, structure, and performance}.
\newblock Asm International, 2015.

\bibitem{nishiyama2012martensitic}
Zenji Nishiyama.
\newblock {\em Martensitic transformation}.
\newblock Elsevier, 2012.

\bibitem{sandvik1983characteristicsI}
BPJ Sandvik and CM~Wayman.
\newblock Characteristics of lath martensite: Part i. crystallographic and
  substructural features.
\newblock {\em Metallurgical transactions A}, 14(4):809--822, 1983.

\bibitem{kelly1992crystallography}
Patrick~M Kelly.
\newblock Crystallography of lath martensite in steels.
\newblock {\em Materials Transactions, JIM}, 33(3):235--242, 1992.

\bibitem{christian1995deformation}
John~Wyrill Christian and Subhash Mahajan.
\newblock Deformation twinning.
\newblock {\em Progress in materials science}, 39(1-2):1--157, 1995.

\bibitem{monzen1992face}
Ryoichi Monzen and Masaharu Kato.
\newblock Face-centred cubic to body-centred cubic martensitic transformation
  of fe-co particles in a copper matrix.
\newblock {\em Journal of materials science letters}, 11(1):56--58, 1992.

\bibitem{le1993effects}
Tuyen~D Le, IM~Bernstein, and S~Mahajan.
\newblock Effects of hydrogen on micro-twinning in a {FeTiC} alloy.
\newblock {\em Acta metallurgica et materialia}, 41(12):3363--3379, 1993.

\bibitem{bilby1965theory}
Bruce~Alexander Bilby and AG~Crocker.
\newblock The theory of the crystallography of deformation twinning.
\newblock {\em Proceedings of the Royal Society of London. Series A.
  Mathematical and Physical Sciences}, 288(1413):240--255, 1965.

\bibitem{vitek1970core}
V~Vitek, RC~Perrin, and DK~Bowen.
\newblock The core structure of $1/2 \langle 111\rangle$ screw dislocations in
  bcc crystals.
\newblock {\em Philosophical Magazine}, 21(173):1049--1073, 1970.

\bibitem{bristowe1976zonal}
PD~Bristowe and AG~Crocker.
\newblock Zonal twinning dislocations in body centred cubic crystals.
\newblock {\em Philosophical Magazine}, 33(2):357--362, 1976.

\bibitem{bristowe1977computer}
PD~Bristowe and AG~Crocker.
\newblock A computer simulation study of the structure of twinning dislocations
  in body centred cubic metals.
\newblock {\em Acta Metallurgica}, 25(11):1363--1371, 1977.

\bibitem{shi2016competing}
Zhe Shi and Chandra~Veer Singh.
\newblock Competing twinning mechanisms in body-centered cubic metallic
  nanowires.
\newblock {\em Scripta Materialia}, 113:214--217, 2016.

\bibitem{rowlands1970application}
PC~Rowlands, EO~Fearon, and M~Bevis.
\newblock The application of the kossel technique and electron microscopy to
  the study of the microstructure of {Fe-32\%} ni martensite crystals.
\newblock {\em Journal of Materials Science}, 5(9):769--776, 1970.

\bibitem{dai2020coherent}
Fu-Zhi Dai, Zhi-Peng Sun, and Wen-Zheng Zhang.
\newblock From coherent to semicoherent—evolution of precipitation
  crystallography in an fcc/bcc system.
\newblock {\em Acta Materialia}, 186:124--132, 2020.

\bibitem{khachaturyan1991adaptive}
AG~Khachaturyan, SM~Shapiro, and S~Semenovskaya.
\newblock Adaptive phase formation in martensitic transformation.
\newblock {\em Physical Review B}, 43(13):10832, 1991.

\bibitem{kitahara2006crystallographic}
Hiromoto Kitahara, Rintaro Ueji, Nobuhiro Tsuji, and Yoritoshi Minamino.
\newblock Crystallographic features of lath martensite in low-carbon steel.
\newblock {\em Acta materialia}, 54(5):1279--1288, 2006.

\end{thebibliography}

\appendix

\section{Rotational matrices for KS OR $V_1$ and $V_2$}
\label{app}

Rotational matrix $J_1$ (resp. $J_2$) that transforms the coordinate system of the parent fcc lattice into the bcc lattice with KS OR $V_1$ (resp. $V_2$) reads \cite{kitahara2006crystallographic}:

\[
J_1=
\begin{pmatrix}
0.742  & -0.667 & -0.075\\
0.650  & 0.742  & -0.167\\
0.167  & 0.075  & 0.983
\end{pmatrix}, \quad
J_2=
\begin{pmatrix}
0.075  & 0.667 & -0.742\\
-0.167  & 0.742  & 0.650\\
0.983 & 0.075  & 0.167
\end{pmatrix}
\]

\section{Experimental conditions}
\label{exp}

In this work, the microstructural evolution during casting of a Cu-2.0Fe-0.5Co (wt. \%) alloy was investigated via transmission electron microscopy (TEM). The alloy was prepared from high purity Cu, Fe and Co (purity of 99.99, 99.50, 99.95 wt. \%, respectively) and elaborated by gravity casting in a vacuum chamber with a medium frequency electrical furnace. A 50 mm $\times$ 65 mm $\times$ 195 mm cuboid specimen was cast after a homogenization at 1300$^{\circ}$C for 20 minutes, followed by an isothermal holding at 1150-1200$^{\circ}$C. Thin TEM foils were prepared from 3 mm diameter discs that were mechanically polished and electronically thinned by low-energy Ar milling. 

TEM and HRTEM characterization was carried out under JEM-2100F Field Emission Electron Microscope. The TEM and HRTEM images were further processed by DigitalMicrograph 3.5 and Gatan Microscopy Suite 2.1 software. The thickness of twin domains within Fe-rich precipitates were measured and determined from the TEM and HRTEM images, and at least 100 twin domains were analyzed for quantification.

\section*{ACKNOWLEDGMENTS}

Part of this work was performed using computing resources of CRIANN (Normandy, France) where simulations were performed as Project No. 2012008. This work was supported by the Agence Nationale de la Recherche (contract C-TRAM ANR-18-CE92-0021), the Beijing Municipal Natural Science Foundation (No. 2214072), and the China Postdoctoral Science Foundation (2019M660451).

\section*{AUTHOR CONTRIBUTIONS}

G. Demange, M. Lavrskyi, R. Patte and H. Zapolsky developed the QA model.  G. Demange performed the simulations. K. Chen, J. Hu, X. Chen and Z. Wang  performed the preparation of experimental samples and TEM/HRTEM observations. All authors participated in the redaction of the manuscript.

\section*{COMPETING INTERESTS}

The authors declare no financial and/or non-financial competing interests.

\section*{CORRESPONDING AUTHORS}
Correspondence to Zidong Wang and Gilles Demange.

\end{document}